\documentclass[aps,prd,nofootinbib,twocolumn,superscriptaddress,letterpaper,preprintnumbers,notitlepage]{revtex4-1}
\usepackage{amsmath, amssymb,braket}
\usepackage{array}
\usepackage{color}
\usepackage{float}
\usepackage{graphicx}
\usepackage{subfigure}
\usepackage{natbib}
\usepackage{adjustbox}
\usepackage[colorlinks=true,citecolor=blue,urlcolor=blue]{hyperref}
\usepackage{circuitikz}
\usepackage{xspace}
\usepackage{mathrsfs}

\newcommand{\dtau}{\Delta\tau_{\rm burst}}
\newcommand{\zx}{z_x}
\newcommand{\sx}{\sigma_x}
\newcommand{\ax}{A_x}
\newcommand{\xe}{x_e}
\newcommand{\arec}{\alpha_{\rm rec}}
\newcommand{\mnu}{\sum m_\nu}
\newcommand{\ev}{\,{\rm eV}}
\newcommand{\eV}{\,{\rm eV}}
\newcommand{\lcdm}{$\Lambda$CDM\xspace}
\newcommand\orcid[1]{\href{http://orcid.org/#1}{\adjustbox{trim={-.15\width} {0\height} {-.15\width} {0\height},clip}{\includegraphics[height=10pt]{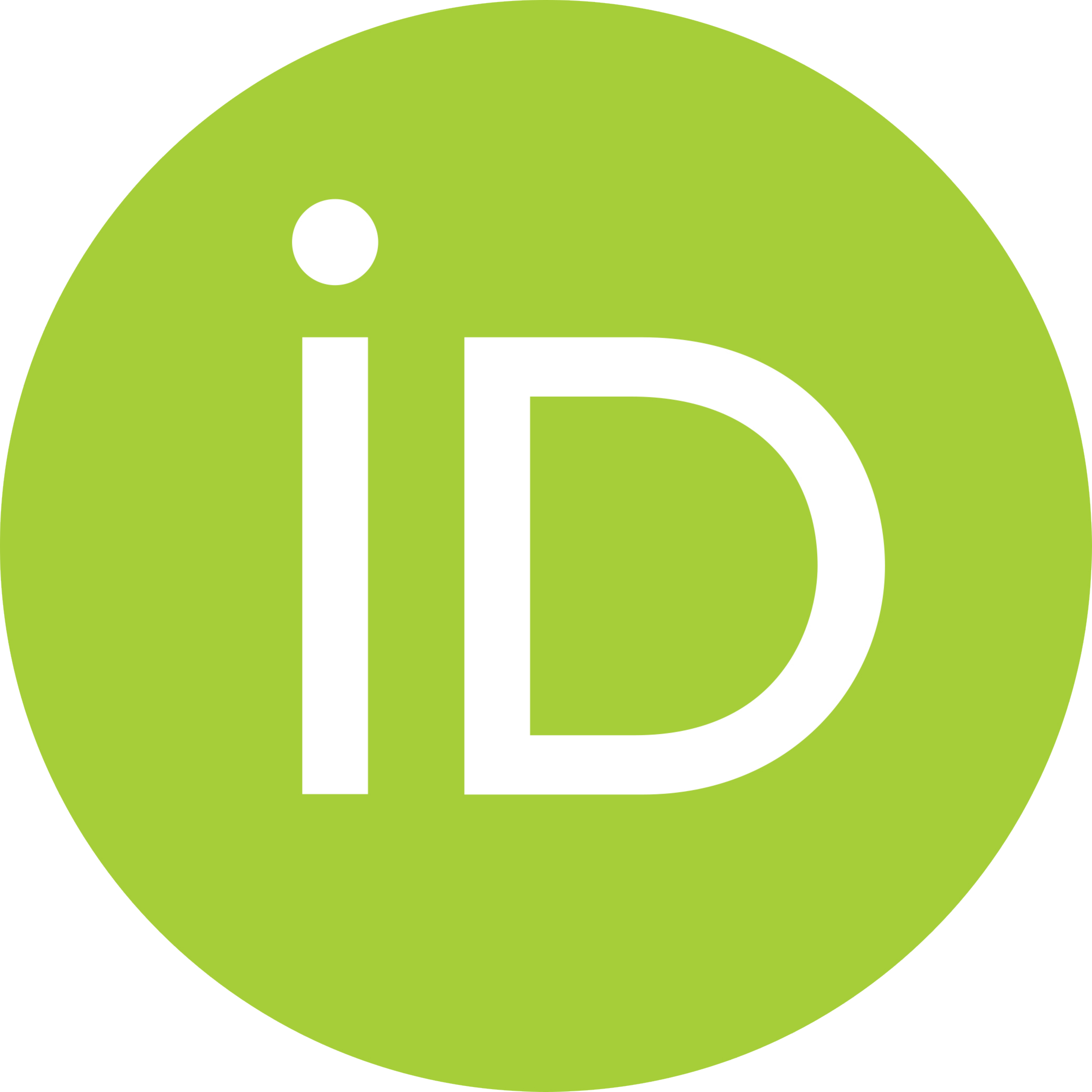   }}}}

\newcommand{\ess}{\emph{s}}

\definecolor{darkblue}{rgb}{0,0,0.55}%
\definecolor{darkgreen}{rgb}{0,0.45,0}%
\newcommand{\txt}[1]{#1}

\newcommand{\eqd}[1]{#1}
\newcommand{\del}[1]{}

\graphicspath{{figures/}{./}}

\begin{document}
\raggedbottom

\title{A Burst of Reionization in the Dark Ages}

\author{Neal Weiner\orcid{0000-0003-2122-6511}}
\affiliation{Center for Cosmology and Particle Physics, Department of Physics, New York University, New York, NY 10003, USA}

\author{Soubhik Kumar \orcid{0000-0001-6924-3375}}\affiliation{Institute of Cosmology, Department of Physics and Astronomy, Tufts University, Medford, MA
02155, USA}
\author{Wenzer Qin\orcid{0000-0001-7849-6585}}\affiliation{Center for Cosmology and Particle Physics, Department of Physics, New York University, New York, NY 10003, USA}
\author{Zilu Zhou\orcid{0000-0002-2800-958X}}\affiliation{Center for Cosmology and Particle Physics, Department of Physics, New York University, New York, NY 10003, USA}
\author{Yacine Ali-Ha\"imoud\orcid{0000-0003-0533-0319}}\affiliation{Center for Cosmology and Particle Physics, Department of Physics, New York University, New York, NY 10003, USA}
\author{Hector Afonso G. Cruz\orcid{0000-0002-1775-3602}}\affiliation{Center for Cosmology and Particle Physics, Department of Physics, New York University, New York, NY 10003, USA}
\date{\today}

\begin{abstract}
Primordial overdensities at small scales can be significantly larger than those observed at large scales. As a result, they can collapse much earlier, at redshifts $z \sim 100$. In these collapsed halos, stars and black holes can form, where they can then contribute to an early reionization of the universe. If this epoch is temporary, the free electrons will recombine yielding a low level of ionization leading into cosmic dawn. This contributes to the optical depth of the universe, and adds more weakly to low-$\ell$ polarization signals of ionization compared to later sources.  We study the effects of a redshift-localized source of reionization $\dtau$ in the context of a broad suite of cosmological data to constrain this scenario, finding a preference of approximately $\sim 2-2.5\sigma$ for a new contribution. Regardless of Planck reprocessing used (PR3 - SRoll2+\textsc{plik} or PR4 - LoLLiPoP+CamSpec/HiLLiPoP), the results are consistent, favoring a new contribution $\dtau \sim 0.012$ ($0.001 < \dtau < 0.022$ at 95\%) and a similar total optical depth $\tau \sim 0.073$. The precise era of reionization is somewhat discrepant depending on the higher-$\ell$ reprocessing used, with HiLLiPoP favoring $z\sim 45$, CamSpec $z\sim 100$ and \textsc{plik} $z\sim 165$.  If such a source is present at $40\lesssim z\lesssim150$, LiteBIRD should be able to detect it, while Simons Observatory should also constrain it.
\end{abstract}

\maketitle

\section{Introduction}
Modern cosmology has settled into a simple and highly effective framework. Tiny adiabatic perturbations, sourced by inflation, with a nearly scale-invariant spectrum, grow upon horizon entry until they become nonlinear and collapse into the galaxies we see today. The evidence for this is strong, borne out by the evolution of structure we see as well as the temperature perturbations in the cosmic microwave background (CMB).  Importantly, it is also supported by the polarization signals of the CMB, i.e., both the optical depth of the  universe and the low-$\ell$ polarization signal~\cite{Planck:2018vyg}, both of which are consistent with a reionization of the universe from early stars at redshifts $z \sim 10\text{--}20$.

However, not all is an unambiguous success. There are well known cosmic anomalies related to the expansion rate of the universe~\cite{Riess:2021jrx}, the amplitude of structure~\cite{Abdalla:2022yfr}, and the time evolution of dark energy in the late universe~\cite{DESI:2025zgx}. In addition, there are signs of large galaxies appearing earlier than expected~\cite{Labbe:2022ahb},
while the origins of supermassive black holes (SMBHs) and so-called ``little red dots (LRDs)~\cite{Matthee:2023utn},'' remain elusive. Even the measurement of the reionization optical depth $\tau$ itself shows a tension: the value inferred directly from the large-scale (low-$\ell$) CMB polarization sits markedly lower than the value preferred once the CMB damping tail and lensing are combined with DESI BAO~\cite{Sailer:2025lxj}, a discrepancy that has been tied to a ``negative neutrino mass'' arising as a fit in cosmological datasets~\cite{Craig:2024tky,Green:2024xbb,Jhaveri:2025neg}.%

As a part of questioning the basic assumptions of the standard cosmological model, one possibility is that at small enough scales, the simple scale-invariant story may not persist. Rare Poisson processes in inflation~\cite{Kim:2021ida} can yield overdensities, for instance, which might produce these rare galaxies~\cite{Kumar:2025gon}. Pushing the limits further, an increase in the power spectrum at $k\gtrsim 10^2 \, h\,{\rm Mpc}^{-1}$ may produce large black hole seeds through direct collapse at $z \gtrsim 200$, possibly providing  an explanation of SMBHs and LRDs~\cite{Qin:2025ymc}.

If such collapse of small scale structure at early times occurs, it is worth considering whether an additional reionization signal is present. This may come from the x-rays of early black hole accretion, high mass x-ray binaries formed from very early Pop III star formation \citep{Xu:2014aja, Sazonov:2016vac}, or perhaps a population of early stars themselves. Regardless, new collapsed structures, forming after $z\sim100$, might source reionization at some time well before cosmic dawn. Irrespective of the source, the question is simple: to what extent do current data constrain the existence of this early, potentially very early, reionization signal?

Recently, groups have begun more earnestly to consider modifications to reionization. ``Flash'' reionization models~\cite{Aggarwal:2026ogm,Tan:2025obi}%
, which are powered by early stars altered by dark matter annihilation and source additional ionization right before cosmic dawn, provide one such proposal. Modifications to reionization can have important implications beyond the cosmic history of energy injection. A number of studies parameterizing a sub-minimal or ``negative'' effective neutrino mass have found that current cosmological data prefer one or the other at the $2\sigma$ level~\cite{Craig:2024tky,Green:2024xbb}. A number of groups have since argued that effectively relaxing the low $\tau$ inferred from large-scale CMB polarization may have important implications for this and other cosmic anomalies~\cite{Jhaveri:2025neg,Sailer:2025lxj,Sullivan:2026tas,Allali:2025yvp}. In light of this, we pose the question more pointedly: what do current data say about a redshift-localized contribution to $\tau$, arising at essentially any era after recombination. We will not attempt to tie ourselves to any particular model or prejudice of what a natural cosmic history ``should'' look like, but only what {\em could} it look like. While our specific motivation arises from the not-quite primordial black hole (NQPBH) scenario of
~\cite{Qin:2025ymc}, we consider that a jumping off point. Completely orthogonal scenarios, such as a pole in dark matter annihilation, or resonant conversion and decay, or structures formed from dissipative dark matter could in principle provide a similar signal. It is precisely not to attach to any specific scenario that we proceed in a model-agnostic fashion.

In this paper we proceed as follows: in Sec.~\ref{sec:model} we lay out the phenomenological parameterization for reionization we consider, and then we proceed by studying this in the context of current cosmological data, finding constraints and allowed regions. In Sec.~\ref{sec:andneutrinos} and~\ref{sec:previous}, we discuss the interplay between neutrino mass and this additional source of $\tau$, and find that, as previously suggested, an additional source of $\tau$ may ameliorate the ``negative neutrino mass tension.'' Finally, we conclude in Sec.~\ref{sec:disc}.%

\section{A phenomenological model and fits}\label{sec:model}

We model an early episode of enhanced ionization as a localized,
sourced perturbation to the recombination history, or a ``burst''. 
We parameterize ionization in terms of the global free electron fraction, $\xe \equiv n_e / n_H$, where $n_e$ is the number density of free electrons and $n_H$ is the number density of all hydrogen.
The perturbation is
described by three parameters: an amplitude $\ax$, a central redshift $\zx$, and a
width $\sx$, which we fix throughout to $\sx=0.1$. Writing
\[
  \ess(z) \equiv \ln(1+z) - \ln(1+\zx),
\]
the ionization excess this parameterization would trace out, were it
imposed directly, is the Gaussian $\ax\,e^{-\ess(z)^2/(2\sx^2)}$. We do not impose
this profile on $\xe(z)$ directly; instead we use its source (i.e., its derivative):
\begin{equation}
\left.\frac{d\xe}{dz}\right|_{\rm source} =
\begin{cases}
  -\dfrac{\ax\,\ess(z)}{\sx^2(1+z)}\,e^{-\ess(z)^2/(2\sx^2)}, & z > \zx,\\[8pt]
  0, & z \le \zx.
\end{cases}
\label{eq:source}
\end{equation}
We add this source term directly into the same equation that \textsc{class}'s~\cite{Blas:2011rf}
thermodynamics module already solves to track the standard recombination
history, which \textsc{HyRec} \cite{Ali-Haimoud_2010, Ali-Haimoud:2010hou} integrates in redshift to
evolve the free-electron fraction $\xe$.

In this form, the signal comes from a new source at $z>\zx$, after which conventional recombination processes take over. The subsequent decline is
governed by \textsc{HyRec}'s unmodified rate equations once the source
switches off---consequently,
the peak of $\xe(z)$ sits slightly
before $\zx$ (by $4$--$8\%$ at the best-fit amplitudes, and less for smaller $\ax$), since the
injection rate peaks at $\ess=\sx$ rather than at $\ess=0$ where it terminates.

\begin{figure}[htbp]
\centering
\eqd{\includegraphics[width=0.5\textwidth]{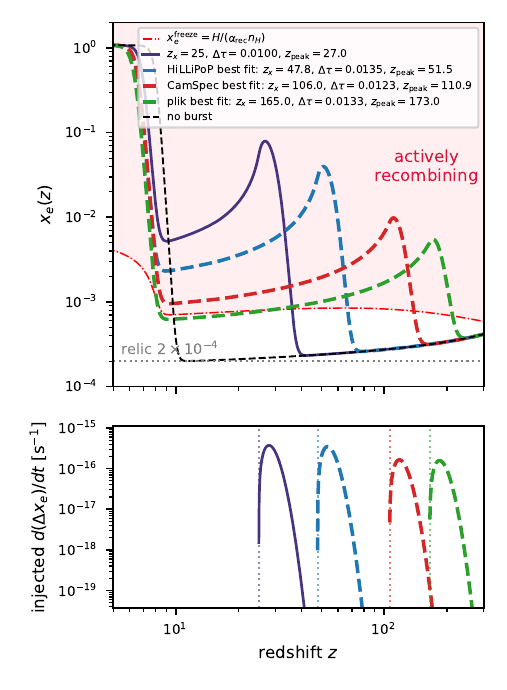}}
\caption{%
    Free-electron fraction $x_e(z)$
    histories for the burst-reionization source of Eq.~\eqref{eq:source}, which switches off at $z_x$ and is followed by unmodified recombination. One curve ($z_x=25$) is illustrative, at a fixed $\dtau=0.010$ (corresponding to $A_x=0.32$); the other three are the actual best-fit points: HiLLiPoP's best fit ($z_x=47.8$, $\ax=0.116$, $\dtau=0.0135$), CamSpec's best fit ($z_x=106$, $\ax=0.0169$, $\dtau=0.0123$), and \textsc{plik}'s best fit ($z_x=165$, $\ax=0.0083$, $\dtau=0.0133$) (Table~\ref{tab:bestfit}). 
    The dash-dotted curve is the freeze-out floor $\xe^{\rm freeze}$ of
    Eq.~\eqref{eq:freeze}.
    The dotted curve marking $x_e = 2 \times 10^{-4}$ shows the floor in a $\Lambda$CDM cosmology.
}%

\label{fig:xebumps}
\end{figure}

The decline in ionization does not return $\xe$ to its unperturbed value. Hydrogen
recombines on a timescale $1/(\arec n_H \xe)$, and once that exceeds the
Hubble time the remaining ionization is effectively frozen in. Equating the
two defines a floor,
\begin{equation}
\xe^{\rm freeze}=\frac{H}{\arec n_H},
\label{eq:freeze}
\end{equation}
where $H$ is the Hubble parameter, and each burst relaxes only as far as this curve, which is what leaves a
residual $\xe$ visible at low $z$. We use
$\arec \simeq 2.59\times10^{-13}\,(T/10^4\,{\rm K})^{-0.833}\ {\rm cm^3\,s^{-1}}$
\citep{OsterbrockFerland2006}, evaluated at the baryon temperature of the
unperturbed history. Some examples of cosmic $x_e$ histories are shown in
Fig. \ref{fig:xebumps}.

\subsection{Implications of Current Data}
\label{sec:implications}
To study these scenarios, we perform a full MCMC analysis of $\Lambda$CDM extended by
the burst parameters. 
In what follows, we find it useful to use  different definitions of (or contributions to) the optical depth, which we list briefly here for the reader's convenience and discuss in greater depth in Appendix~\ref{sec:tau_definition}:
\begin{itemize}
    \item $\tau_{\rm reio}$ : the value of the optical depth sampled in the \textsc{class} Boltzmann code, which includes contributions between $z=0$ and $z_{\rm min}$, the minimum of $\xe(z)$. This can variably contain $\dtau$, due to the definition of $z_{\rm min}$

    \item $\tau_{\rm eff}$ : the effective total optical depth, including contributions from conventional reionization and exotic sources.

    \item $\dtau$ : the exotic or ``burst" contribution to $\tau_{\rm eff}$.

    \item $\tau_{\rm low\text{-}z}$ : the low-$z$ reionization contribution to $\tau_{\rm eff}$ (by definition, $\tau_{\rm low\text{-}z}= \tau_{\rm eff}-\dtau$).
\end{itemize}

We begin by fixing to a conventional neutrino cosmology with normal neutrino mass ordering, i.e., setting the effective number of neutrino species to $N_{\rm eff}=3.044$ and sum of neutrino masses to $\sum m_\nu = 0.059\ev$.
We sample the six $\Lambda$CDM parameters
$\{\omega_b,\omega_{\rm cdm},H_0,\ln(10^{10}A_s),n_s,\tau_{\rm reio}\}$ , plus the source parameters $(\dtau, z_x)$ at fixed $\sigma_x=0.1$, where $\dtau$ is the contribution to the optical depth from the modeled burst. For reasons that we explain in the Appendix~\ref{sec:tau_definition}, we will report our results in terms of the total optical depth $\tau_{\rm eff}$, rather than $\tau_{\rm reio}$, although $\tau_{\rm eff}$ reduces to $\tau_{\rm reio}$ in most cases.
We define the burst model using the source described previously with $\sigma_x$ fixed.
That is, for $\sigma_x\lesssim 0.2$ the effect on the fit is rather small, and the contribution to $x_e$ is rather sudden. For larger values of $\sigma_x$, the contribution becomes much broader, and the scenario would no longer be properly considered the same physical model (see
Fig.~\ref{fig:xebumps_sigma} in the Appendix). We adopt flat priors on the new contribution to reionization $\dtau\in[0,0.12]$ (which we define more precisely in the appendix) and
$z_x\in[10,300]$. The amplitude $A_x$ is
obtained by inverting a precomputed table in $(\dtau, z_x)$ built at a fixed fiducial cosmology (see Appendix~\ref{sec:mcmc_appendix} for details). The preferred $\dtau$ range does not significantly depend on whether we sample in $\dtau$ or $A_x$, while $z_x$ does.\footnote{Because the same $A_x$ at lower redshifts yields a lower $\dtau$, when imposing physical cutoffs on the amount of $\tau$ allowed, there is a much larger volume at low $z_x$, assuming a flat $A_x$ prior. On the other hand, a log prior in $A_x$ then essentially oversamples $\dtau=0$ at low redshift, skewing our posteriors in the opposite direction. Working with a flat prior in $\dtau$ is a compromise.} 
Theory spectra are computed with \textsc{class}~\cite{Blas:2011rf}.
Nuisance and instrumental calibration parameters intrinsic to the likelihoods (relative
calibration factors, foreground amplitudes, etc.) are sampled alongside the cosmological
and burst parameters and marginalized over in the reported posteriors, as we describe in the appendix.
Posteriors are obtained with \textsc{cobaya}'s~\cite{Torrado:2020dgo} Metropolis--Hastings
sampler using an adaptively learned proposal covariance. To verify that the burst reionization is correctly captured by the recombination and CMB codes, we independently implemented this model in the JAX based Einstein-Boltzmann solver \textsc{abcmb}~\cite{Zhou:2026eit}. This implementation was used along with the cosmology emulator pipeline \textsc{olé}~\cite{Gunther:2025xrq} for running MCMC. We performed separate MCMC chains with the two codes using the likelihoods for producing Fig.~\ref{fig:tau_overlay_lollipoponly} (see below), and obtained posteriors that agreed to well within 1-$\sigma$. Note that \textsc{class} is still the primary Boltzmann solver used in the rest of this work; this independent analysis only serves as a cross-check.

Unless otherwise noted, the CMB and large-scale-structure likelihood combination
consists of: SPT-3G D1 temperature and polarization~\cite{SPT-3G:2025bzu}
(\texttt{candl} implementation~\cite{Balkenhol:2024sbv}, restricted to $\ell>1000$);
ACT DR6 CMB-only power spectra~\cite{ACT:2025fju} (restricted to
$1000<\ell<3500$, so as not to double-count information already carried by
the {\it Planck}-based high-$\ell$ likelihoods below) and CMB
lensing~\cite{ACT:2023dou,ACT:2023kun};  the Pantheon$+$ Type~Ia supernova
compilation~\cite{Scolnic:2021amr,Brout:2022vxf}; low-$\ell$ TT {\it Planck} power spectra~\cite{Planck:2019nip}; and DESI DR2 baryon acoustic
oscillation distance measurements~\cite{DESI:2025zgx}. Additionally we include \emph{Planck} low-$\ell$ $EE$ polarization as well as high-$\ell$ {\it Planck} temperature and polarization ($TT+TE+EE$ throughout). For low-$\ell$ we  use either (a) NPIPE
(PR4)~\cite{Planck:2020olo}  (\textsc{LoLLiPoP}) ~\cite{Tristram:2020wbi,Tristram:2023haj} combined with either 
CamSpec~\cite{Efstathiou:2019mdh,Rosenberg:2022sdy} or 
\textsc{HiLLiPoP}~\cite{Tristram:2023haj} for high-$\ell$ {\it Planck} or (b) PR3 SRoll2 for low-$\ell$ polarization, along with \textsc{plik} for high-$\ell$ {\it Planck} data. 
The full data suites we distinguish as CamSpec+, HiLLiPoP+ or \textsc{plik}+.

\subsection{Results}
\label{sec:results}

We proceed in stages, starting from the single most sensitive existing probe of
this epoch alone, then broadening the dataset, to make clear
which data source the eventual preference for $\dtau > 0$.

\subsubsection{Low-$\ell$ alone}
\label{sec:lollipop_alone}
\emph{Planck}  low-$\ell$ $EE$ polarization (implemented in SRoll2 or \textsc{LoLLiPoP}) is,
on its own, the single most constraining current probe of $\tau$ and its
redshift dependence: it is a  likelihood over exactly the multipole
range ($\ell\lesssim30$) most sensitive to when reionization happened, with no
high-$\ell$ foreground or calibration nuisances to marginalize over. We
therefore start with the question in its narrowest form: what implications do the low-$\ell$ data, alone, have for a burst of reionization, and what constraints can we place?

Low-$\ell$ data alone does not constrain other cosmological parameters, so we begin by fixing to a fiducial cosmology  floating only Planck's absolute calibration nuisance parameter $A_{\rm planck}$, $A_s$, $\tau_{\rm reio}$ and $z_x$ and $\dtau$ in the burst model. When using LoLLiPoP we fix to the joint HiLLiPoP best-fit cosmology in Table~\ref{tab:bestfit}, and when using SRoll2 we fix to the joint \textsc{plik} best-fit cosmology in Table~\ref{tab:bestfit}.

\begin{figure}[H]
\centering
\includegraphics[width=0.48\textwidth]{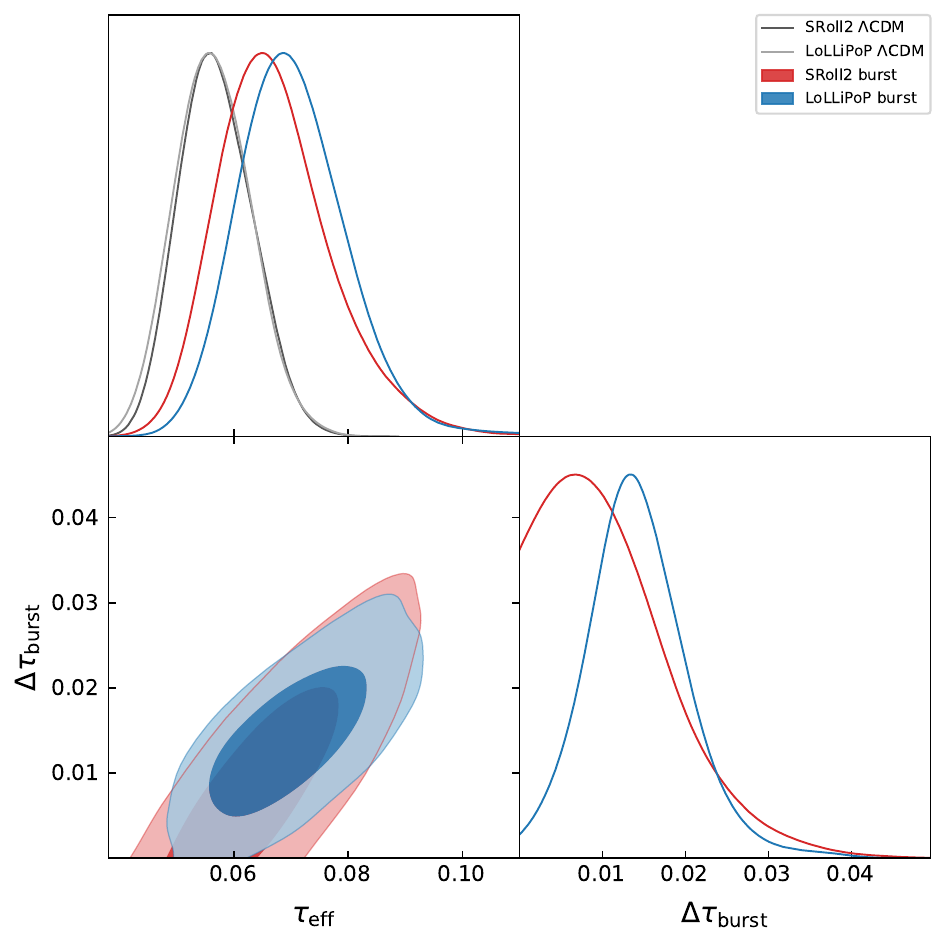}
\caption{Joint posterior on the total optical depth $\tau_{\rm eff}$, the burst
contribution $\dtau$, using only a low-$\ell$ $EE$
likelihood: \textsc{LoLLiPoP} (PR4) and SRoll2 (PR3), each at its own
generation's best-fit cosmology (HiLLiPoP and \textsc{plik} respectively,
Table~\ref{tab:bestfit}), against the matched \lcdm\ control for each.}
\label{fig:tau_overlay_lollipoponly}
\end{figure}

At the fiducial cosmologies, compared to \lcdm\ (both refined by the same
iterative minimization, see Sec.~\ref{sec:mcmc_appendix}), using LoLLiPoP we find a preference for the additional source by
$\Delta\chi^2\simeq +4.4$, with a new optical depth contribution of $\dtau\approx0.010$ (see Fig.~\ref{fig:tau_overlay_lollipoponly}) and its era, $z_x$, essentially unconstrained in the range of study. Using SRoll2, in contrast, we find an improvement of only $\Delta\chi^2\simeq +0.4$. The range for $\dtau$ is consistent regardless of dataset.

\subsubsection{Adding the full data suite}
\label{sec:full_suite}
With this result, we then add the remainder of the dataset described in
Sec.~\ref{sec:implications}: {\it Planck} low-$\ell$ TT, ACT DR6, SPT-3G, DESI BAO, and Pantheon$+$,
together with a high-$\ell$ \emph{Planck} likelihood. For PR3, we use \textsc{plik}. With PR4, we must make a
choice, since there are two available high-$\ell$
\emph{Planck} NPIPE likelihoods: CamSpec and HiLLiPoP. We
therefore report both. 
Throughout, the quoted $\Delta\chi^2$ includes the
nuisance-parameter Gaussian priors (Sec.~\ref{sec:mcmc_appendix}); that total
is the quantity every best fit was minimized on, and the data-alone number is
the same two points' data $\chi^2$ difference rather than a separate
optimization.

Regardless of whether we include PR3 or PR4, adding the full suite grows the preference found, and arrives at a largely consistent point, with only some disagreement on the preferred value of $z_x$. 

We find
\begin{itemize}
\item \textbf{SRoll2+\textsc{plik}}: $\Delta\chi^2=6.90$, or $6.10$ counting data alone.
\item \textbf{LoLLiPoP+CamSpec}: $\Delta\chi^2=8.33$, or $8.05$ counting data alone.
\item \textbf{LoLLiPoP+HiLLiPoP}: $\Delta\chi^2=7.29$, or $6.85$ counting data alone.
\end{itemize}
The full suite pushes a new source's significance up from the mild
LoLLiPoP-alone preference, or negligible SRoll2-alone preference.

It is worth understanding where the preference comes from. Partitioning the
total $\Delta \chi^2$ of a best-fit point across datasets is of limited utility: the point is fitted jointly, so the parameters at the minimum already
reflect a compromise among all the likelihoods, and the share attributed to
any one of them is not truly a ``preference'' of that dataset, itself. Thus, we try a different approach:
we study the $\Delta \chi^2$ when we remove a subset of the included datasets
and search for best fit points. We report this in Table~\ref{tab:dropsets}.

\begin{table*}[t]
\centering
\begin{tabular*}{\textwidth}{@{\extracolsep{\fill}}lccc}
\hline
 & \multicolumn{2}{c}{PR4 (NPIPE)} & PR3 \\
Datasets used & CamSpec & HiLLiPoP & \textsc{plik} \\
\hline
full suite                      & $8.33$ & $7.29$ & $6.90$ \\
$-$ DESI BAO                    & $6.44$ $[-1.89]$ & $6.32$ $[-0.97]$ & $3.71$ $[-3.19]$ \\
$-$ ACT DR6 $+$ SPT-3G          & $6.43$ $[-1.90]$ & $5.83$ $[-1.46]$ & $4.50$ $[-2.40]$ \\
$-$ ACT lensing                 & $7.77$ $[-0.56]$ & $7.25$ $[-0.04]$ & $5.82$ $[-1.08]$ \\
$-$ Pantheon$+$                 & $8.22$ $[-0.11]$ & $7.36$ $[+0.07]$ & $6.93$ $[+0.03]$ \\
$-$ DESI $+$ ACT DR6 $+$ SPT-3G & $4.70$ $[-3.63]$ & $4.81$ $[-2.48]$ & $2.63$ $[-4.27]$ \\
$-$ low-$\ell$ $TT$             & $8.54$ $[+0.22]$ & $7.25$ $[-0.04]$ & $6.99$ $[+0.09]$ \\
$-$ high-$\ell$ \emph{Planck}   & \multicolumn{2}{c}{$10.36$ $[+2.03]$ / $[+3.07]$} & $5.10$ $[-1.80]$ \\
$-$ low-$\ell$ $EE$ & $0.30$ $[-8.03]$ & $0.06$ $[-7.23]$ & $0.38$ $[-6.52]$ \\
\hline
\end{tabular*}
\caption{$\Delta\chi^2$ (including the nuisance-parameter Gaussian priors)
between the burst model and its matched \lcdm\ control, when the listed
dataset is removed from the full suite and both models are re-minimized. The columns correspond to CamSpec+, HiLLiPoP+ and \textsc{plik}+}
\label{tab:dropsets}
\end{table*}

The table tells an interesting story. We start with PR4. Outside of LoLLiPoP, the additional support for the burst model comes dominantly from the very high-$\ell$ datasets ACT DR6 and SPT, and DESI. ACT lensing  and Pantheon+ play marginal roles. 

If we remove the LoLLiPoP likelihood, the preference for the burst goes away entirely. However, this is misleading. In doing this, the \lcdm\, fit slides $z_{\rm reio}$ up to $\sim 11.5$ in order to achieve $\tau \simeq 0.095$. Such large values of $\tau$ and $z_{\rm reio}$ are in tension with a number of independent observations (e.g., \cite{Kageura:2026ryq}). Thus, one can remove LoLLiPoP but cap $z_{\rm reio}$ at some conservative value (such as $z_{\rm reio}<10$) and redo the fit. This occurs, however, by the model again setting $z_{\rm reio}$ to the highest allowed value, and using the burst to raise the total $\tau$ back to $0.095$. In short, the fit first saturates $\tau_{\rm low\text{-}z} \equiv \tau_{\rm eff} - \dtau$ and then adds a burst contribution to bring it up. As a consequence, the model is extremely sensitive to any prior on  $\tau_{\rm low\text{-}z}$ and $z_{\rm reio}$, and should not be thought of as a real statement of the role the dataset is playing in the overall preference. In contrast, the points fitting the other ``drop-one" (or ``drop-two") datasets lie at real, otherwise reasonable values of \lcdm parameters. 

The same caveat applies to the PR3 column, where removing SRoll2 collapses the preference from $6.90$ to $0.38$: there, too, the \lcdm\ fit answers by raising $\tau$ to $0.098$, so the entry reflects the loss of the only likelihood anchoring the optical depth rather than a measure of what SRoll2 alone contributes. Indeed, as we have already noted, SRoll2 \emph{on its own} prefers the burst by only $\Delta\chi^2=0.38$, so it is really a tension between the low-$\ell$ $EE$ likelihood and the high-$\ell$ data that creates the preference, not any one dataset.

\begin{figure}[t]
\centering
\includegraphics[width=0.48\textwidth]{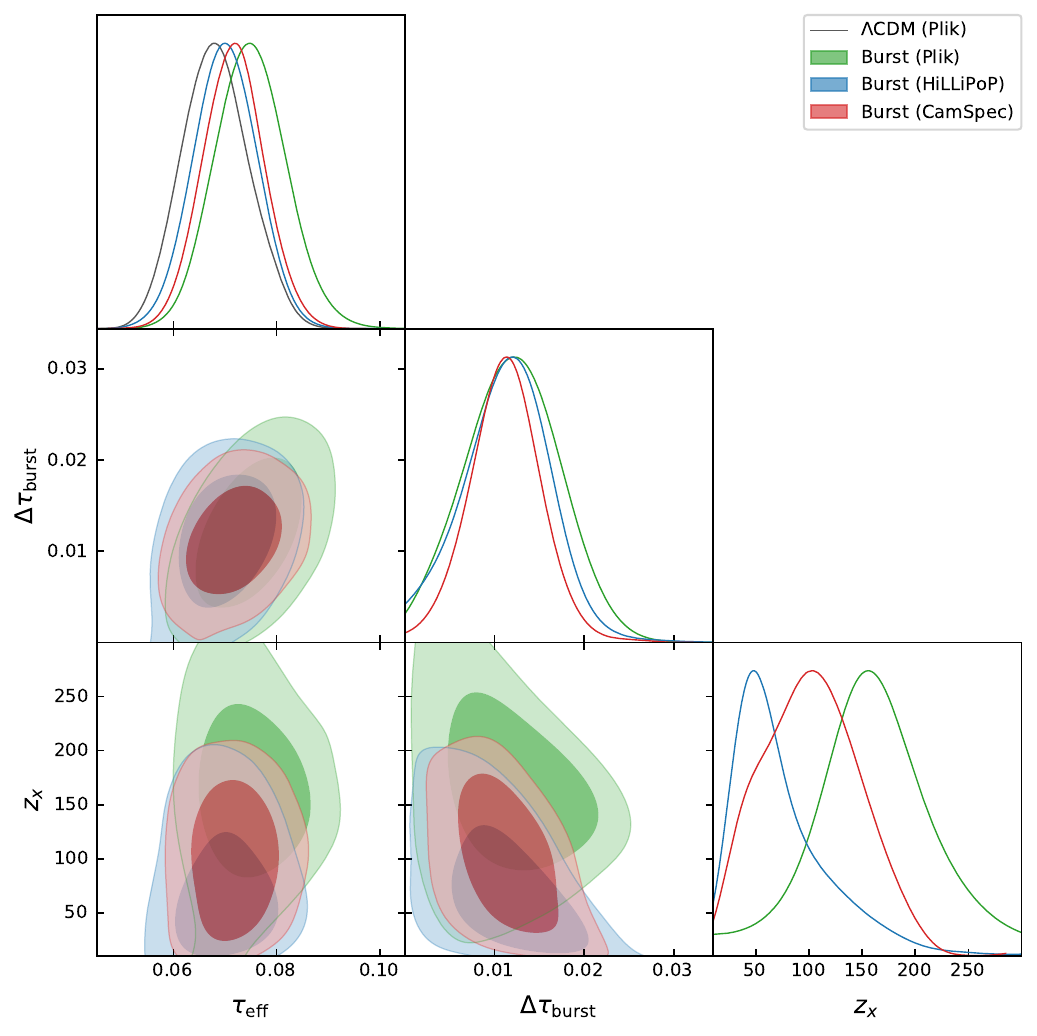}
\caption{Posterior in $\tau_{\rm eff}$, $\dtau$, and $z_x$ for the burst model
under all three \emph{Planck} reprocessings: \textsc{plik} (PR3, green),
CamSpec (PR4, red), and HiLLiPoP (PR4, blue), against the \lcdm\ model
(\textsc{plik}, grey). }
\label{fig:triangle_overlay}
\end{figure}
There is some tension in the PR4 data. The two choices of high-$\ell$ likelihood yield a similar best fit overall $\dtau$ and approximately equal $\Delta \chi^2$, however they  differ quite noticeably in the preferred location
$z_x$. Marginalizing over the full posterior of each run (same dataset and
priors otherwise, only the high-$\ell$ likelihood swapped), we find the two
place their weight quite differently, with median $z_x=102$ for CamSpec
against $z_x=62$ for HiLLiPoP; the 95\% ranges are given in
Table~\ref{tab:bestfit}. Fig.~\ref{fig:triangle_overlay}
shows the full $\tau_{\rm eff}$--$\dtau$--$z_x$ posterior for both choices,
together with \lcdm.

\begin{figure}[t]
\centering
\includegraphics[width=0.48\textwidth]{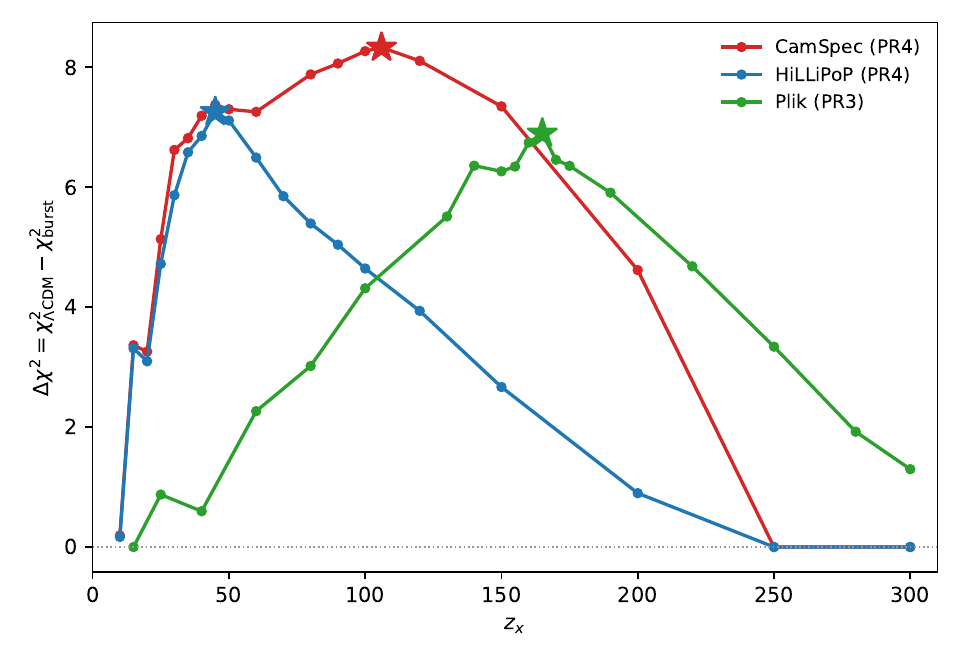}
\caption{Profile likelihood in $z_x$. At each grid value $z_x$ is held fixed
and every other parameter, i.e., six \lcdm parameters, $\ax$, and the full floating nuisance set, is re-minimized.
Plotted is
$\Delta \chi^2=\chi^2_{\Lambda{\rm CDM}}-\chi^2_{\rm burst}$, total and
prior-inclusive, against the matched \lcdm\ control for each reprocessing.
Stars mark the deepest cell on each curve.}
\label{fig:zx_profile}
\end{figure}

This is not an artifact of posteriors. We perform a followup profile-likelihood analysis to understand $z_x$ for all reprocessings, shown in Fig.~\ref{fig:zx_profile}, and see there is a genuine difference in the preference of HiLLiPoP versus either CamSpec or \textsc{plik}. 

\begin{figure*}[p]
    \centering
    \includegraphics[width=0.88\textwidth]{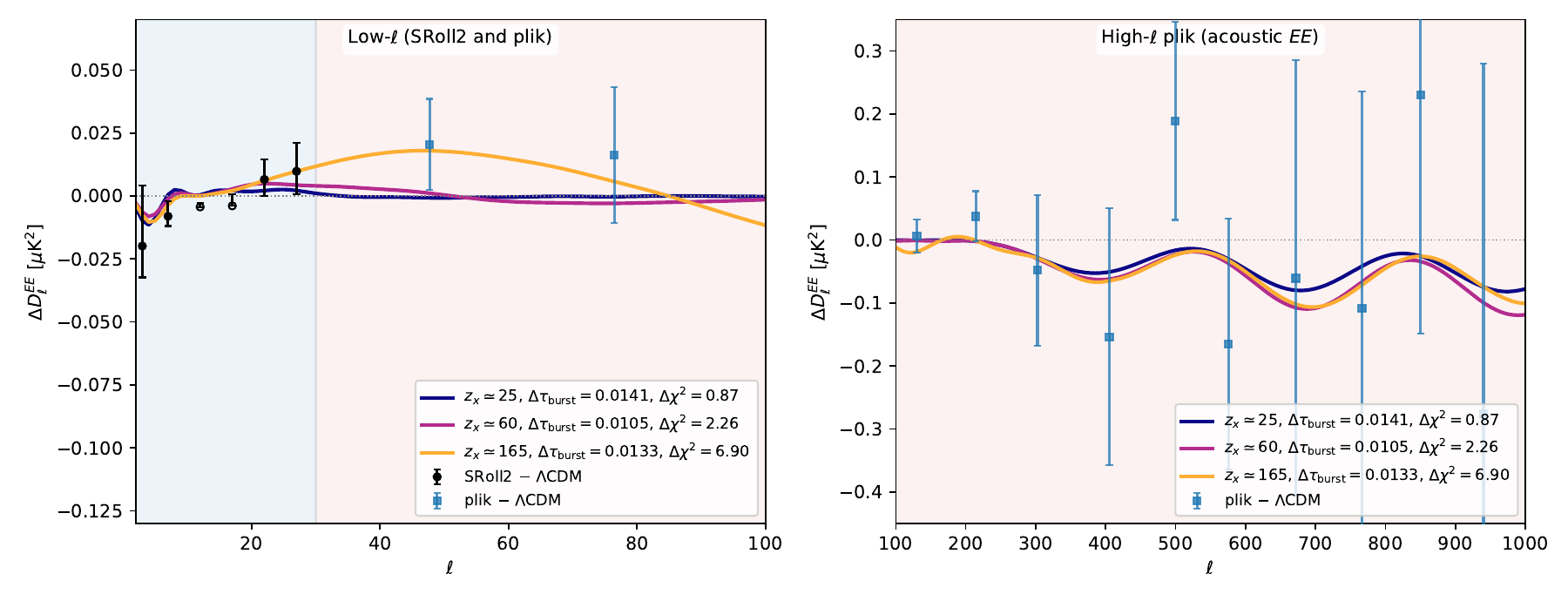}
    \caption{$EE$ residual (burst $-$ $\Lambda$CDM) vs.\ $\ell$ for the
    \textsc{plik}$+$SRoll2 joint fit, at dedicated $\chi^2$-minimized points
    $z_x=25,60,165$, against the low-$\ell$ and high-$\ell$ $EE$ data. Both panels are rebinned
    using the data's own covariance (see text). Circles are SRoll2, squares
    \textsc{plik}.}
    \label{fig:pr3residual}
\end{figure*}

\begin{figure*}[p]
    \centering
    \includegraphics[width=0.88\textwidth]{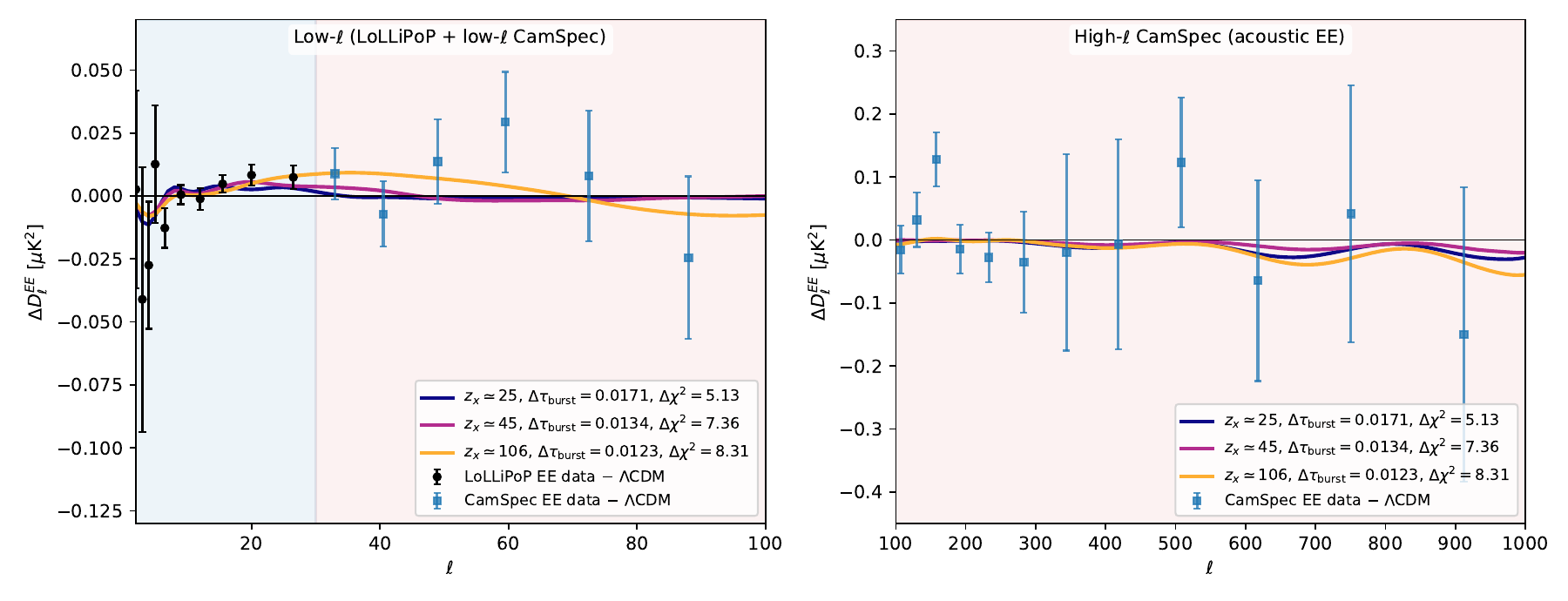}
    \caption{Same as Fig.~\ref{fig:pr3residual}, for the CamSpec-based joint fit
    instead, at its own $\chi^2$-minimized best-fit points at $z_x=25,45,106$.}
    \label{fig:hilresidual_camspec}
\end{figure*}

\begin{figure*}[p]
    \centering
    \includegraphics[width=0.88\textwidth]{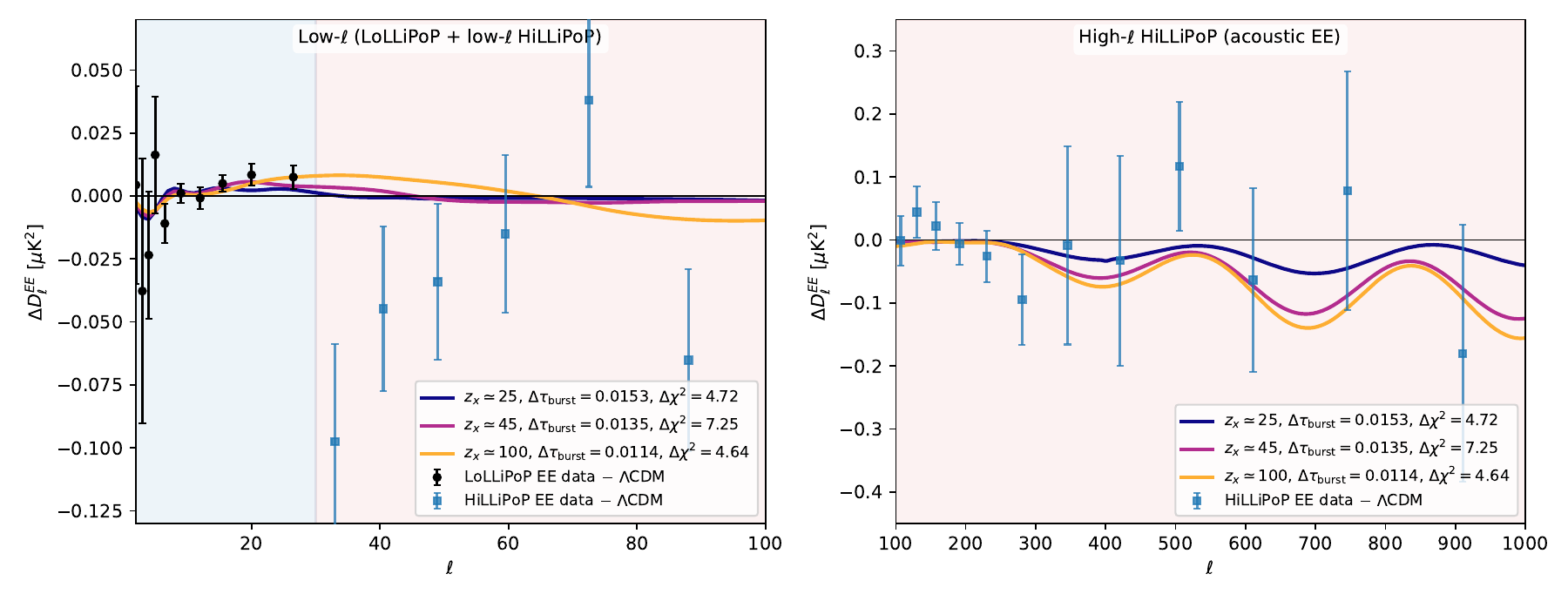}
    \caption{Same as Fig.~\ref{fig:pr3residual}, for the HiLLiPoP-based
    joint fit instead, at its own $\chi^2$-minimized best-fit points at
    $z_x=25,45,100$.}
    \label{fig:hilresidual_hillipop}
\end{figure*}
We can understand this by looking at the shape of the polarization power spectra in Figs.~\ref{fig:pr3residual}, \ref{fig:hilresidual_camspec} and~\ref{fig:hilresidual_hillipop}. In these plots we have rebinned the data to more clearly illustrate trends without scatter, as we describe in Appendix \ref{sec:rebinning}. This is just illustrative - in all our analyses we use full likelihoods.

In the low-$z_x$ range, the CamSpec+LoLLiPoP and HiLLiPoP+LoLLiPoP and even SRoll2+\textsc{plik} agree well (as they should, since the burst does not change the $30< \ell <100$ directly for sufficiently small $z_x$). However, at higher $z_x$, the signal of the burst moves into this intermediate $\ell$ regime. HiLLiPoP begins disfavoring it almost immediately, while CamSpec and \textsc{plik} do not, and peak higher. 
The origin of this discrepancy traces to a
tension in HiLLiPoP's own data. HiLLiPoP's $EE$ bins in the range
$30\lesssim\ell\lesssim100$ fit the \lcdm\ best-fit point noticeably worse
than the rest of HiLLiPoP's own multipole range
(in our fit $\chi^2/{\rm dof}\simeq1.5$ for that stretch alone vs.\ $\sim 1.10$ elsewhere). Including an additional source stresses this further, leading to particularly stringent limits on the allowed range of $z_x$.

In contrast, CamSpec shows no such tension in the data it retains, and as a consequence the allowed $z_x$ range is much broader. This likely traces to CamSpec's own multipole-range choices, which follow the  methodology of its original construction paper~\cite{Efstathiou:2019mdh}, applied here to the PR4 data release we use~\cite{Rosenberg:2022sdy}. Without delving into details, we note that the original PR3-era analysis found at $\ell\lesssim50$ ``clear evidence of systematics'' in spectra involving $217$~GHz~\cite{Efstathiou:2019mdh}, and noted that including that data has ``negligible impact\ldots for $\Lambda$CDM-like cosmologies,'' but produces ``inconsistencies between the $TT$ and $EE$ solutions for models with low frequency oscillatory features in the primordial power spectrum.'' So CamSpec's broader allowed $z_x$ range may partly reflect its more aggressive removal of the same problematic low-$\ell$ $EE$ data, rather than a genuine absence of the underlying tension. Thus, for a \lcdm analysis, both reprocessings should be robust, since $\tau$ is dominated by LoLLiPoP, and the cosmological parameters are dominated by the broader multipole features. In contrast, in this burst model, this region is particularly critical for determining the shape, and any issues of contamination are especially important. It is beyond the scope of our work to discuss the merits of one reprocessing over another, but merely to highlight the sensitivity of the burst scenario to this $\ell$ region.

Thus, we focus on what appears robust between the two choices: namely, a moderate preference ($\sim 2 -  2.5 \sigma$  or $\Delta \chi^2 \sim 8$ for 2 dof) for $\dtau \simeq 0.012$ and an overall increase to $\tau \simeq 0.073$. While the significance is too small to make dramatic claims, it is interesting that the data not only allow, but prefer, an additional, localized source of reionization in the dark ages, which, moreover, raises the total optical depth above \lcdm  without the burst.

\section{Neutrino Mass}%
\label{sec:andneutrinos}%
One important goal of the program of modern cosmology is to measure the imprint of neutrino mass. Terrestrial experiments have precisely measured the mass splittings $\Delta m^2$ amongst the neutrinos, providing a concrete target of $0.059\,$eV in a normally ordered hierarchy ($0.1\,$eV in an inverted hierarchy). The results of the recent DESI mission have complicated this, providing strong constraints~\cite{DESI:2025zgx}, nearly excluding the measured neutrino mass. Efforts to understand this tension have parameterized ``negative neutrino mass'' to explore the full posterior and found that indeed, the combined data do  prefer a negative neutrino mass at the level of about $2 \sigma$~\cite{Craig:2024tky,Green:2024xbb}, sharpening to a $2.8$--$3.3\sigma$ tension when compared directly against the minimum mass allowed by oscillation experiments~\cite{Elbers:2024sha} (see also~\cite{Graham:2025dqn,Herold:2024nvk,Loverde:2024nfi} for related model-building and cosmological measurements of neutrino mass).

It has been argued that additional contributions to $\tau$, by allowing a larger $A_s$, could ameliorate this problem~\cite{Jhaveri:2025neg,Sailer:2025lxj,Sullivan:2026tas}, and so we pursue that study here. We thus  expand our MCMC search to allow variable neutrino masses.

\begin{figure}[htbp]
\centering
\eqd{\includegraphics[width=0.5\textwidth]{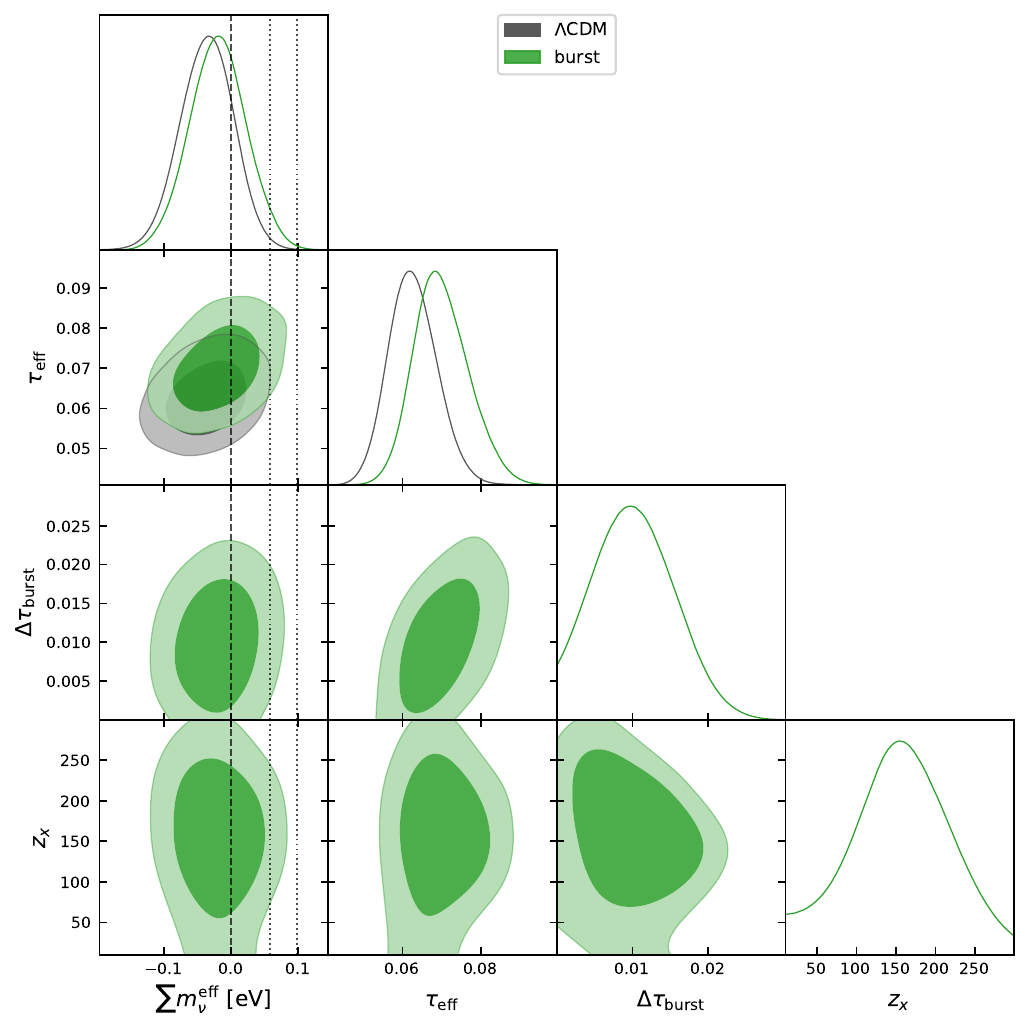}}%
\caption{Triangle plot (1D marginals plus the joint 68\%/95\% contour using the 
\textsc{plik}+ data suite) in
$(\sum m_\nu^{\rm eff},\tau_{\rm eff},\dtau,z_x)$ for \lcdm\ (grey) versus the
burst model (green), both with effective neutrino mass floated (analytically
continued to negative values). 
The dashed line marks
$\sum m_\nu^{\rm eff}=0$; the dotted lines mark the minimal $\sum m_\nu$
allowed by oscillations, $0.059\ev$ (normal ordering) and $0.099\ev$
(inverted).
 }
\label{fig:bump_negmnu_triangle}
\end{figure}

We include neutrino masses in the context of a concrete model, using the full
production dataset with the {\it Planck}~2018 reprocessing (SPT-3G + ACT~DR6 + {\it Planck} \textsc{plik} $TTTEEE$ + Commander low-$\ell$ $TT$ + SRoll2 low-$\ell$ $EE$ + Pantheon+ + DESI~DR2 BAO), following
the prescription of Refs.~\cite{Craig:2024tky,Green:2024xbb}. To study the effect of the burst, we let the burst parameters float, i.e., $\dtau$ and $z_x$ are sampled alongside all \lcdm\ parameters, including the optical-depth parameter $\tau_{\rm reio}$, and thus the neutrino result does not rest on having pinned the source at one redshift. We show the results in
Fig.~\ref{fig:bump_negmnu_triangle}.

The result is a shift in the optical depth and sum of neutrino masses. The posterior median shifts from $-0.035\ev$ in \lcdm\ to $-0.021\ev$ in the burst model. The posterior at the minimum neutrino mass sum $\sum m_\nu^{\rm eff}=0.059 \eV$ is improved: $P_{\rm burst}(0.059\,{\rm eV})/P_{\Lambda{\rm CDM}}(0.059\,{\rm eV})=3.6$, and $\sum m_\nu^{\rm eff}>0$, the physically allowed region, has $18\%$ posterior weight in \lcdm\ and $30\%$ in the burst model, ameliorating, but not eliminating, the tension. The largest effect is from the enhanced value of $\tau$, as has been discussed; our scenario simply provides a means consistent with low-$\ell$ data to increase it.

\section{Relation to prior work}\label{sec:previous}
The possibility that a larger reionization optical depth can relieve the tension between DESI baryon acoustic oscillations and the CMB---which within $\Lambda$CDM manifests as a preference for a sub-minimal or ``negative'' neutrino mass~\cite{Craig:2024tky,Green:2024xbb,DESI:2025zgx}---has been examined from several directions. It has been shown that raising $\tau$ toward $\simeq0.09$ removes much of this tension~\cite{Jhaveri:2025neg,Sailer:2025lxj}, and a higher $\tau\simeq0.11$ has more recently been argued to simultaneously reconcile the neutrino-mass, $H_0$, and dark-energy tensions~\cite{Sullivan:2026tas}; the correlations between $\tau$ and these anomalies have been systematized in Ref.~\cite{Allali:2025yvp}. 

A specific high-redshift source has already been tested against the same discrepancy: Hawking radiation from evaporating primordial black holes, which supplies at most $\dtau\simeq0.008$ and which the CMB data do not significantly prefer~\cite{Sierra:2026vtz}. Whether the same increase can instead be mimicked by a suppression of large-scale primordial power~\cite{Huang:2025xyf} or by inflationary features imprinted directly in the CMB~\cite{Jhaveri:2026qzi} has also been examined. Where an explicit ionizing source has been proposed, it has been tied to a specific epoch: the Population~III.1 ``flash''~\cite{Tan:2025obi}, whose consistency with patchy-kSZ limits was examined in~\cite{Aggarwal:2026ogm}, deposits the extra optical depth at $z\simeq20$--$25$. 

Model-independent characterizations of the reionization history from the CMB have likewise been developed, using principal components~\cite{Hu:2003gh,Mortonson:2007hq,Heinrich:2016ojb,Heinrich:2018btc,Obied:2018qdr,Heinrich:2021ufa}, the non-parametric ``FlexKnot'' reconstruction~\cite{Millea:2018bko,Planck:2018vyg}, although these studies generally restricted contributions to reionization to be at $z \lesssim 50$.
Ref.~\cite{Millea:2018bko} bounds the high-redshift contribution at
$\tau(15,30)<0.015$ (95\%), but their reconstruction extends only to $z=30$,
whereas our preferred bursts sit at $\zx\simeq40$--$165$. Integrating our own
best-fit histories over the same window gives $\tau(15,30)=0.0001$--$0.0009$, one
to two orders of magnitude below that bound; even across the full $\zx$ prior it
never exceeds $0.011$. Essentially all of our $\dtau$ is deposited above $z=30$,
where their reconstruction has no support.
More recently, Gaussian-process reconstructions that define and bound a high-redshift contribution to $\tau$~\cite{Cheng:2025cmb}. 
Importantly, Ref.~\cite{Cheng:2025cmb}, in addition to having a number of methodological differences compared to our approach, used the older SimAll rather than SRoll2 or LoLLiPoP in their analysis. 

An attempt to measure $\tau$ independently of large-scale CMB polarization, within $\Lambda$CDM, in fact finds a mild preference for a higher value, $\tau=0.080\pm0.012$, roughly $1.8\sigma$ above the standard low-$\ell$-polarization determination~\cite{Giare:2023ejv}. It is intriguing that this lands at approximately the same value as our own best-fit $\tau_{\rm eff}$ (Table~\ref{tab:bestfit}, $0.0713$--$0.0760$ depending on high-$\ell$ likelihood). The comparison is the appropriate one: Ref.~\cite{Giare:2023ejv} infers $\tau$ from the damping of the acoustic peaks, and $\tau_{\rm eff}$ is the quantity that appears there, whereas a $\tau$ inferred from the low-$\ell$ $EE$ reionization bump constrains $\tau_{\rm low\text{-}z}$ instead (Appendix~\ref{sec:tau_definition}).  It is not clear what, if anything, to make of a $\Lambda$CDM-only analysis of a different dataset combination landing on essentially the same total $\tau$ that we find once a burst component is included, but the coincidence is notable. Newer independent probes of the low-redshift optical depth, such as the Lyman-$\alpha$ forest~\cite{Garcia-Gallego:2025jrn} and the reionization history itself~\cite{Kageura:2026ryq}, offer a complementary handle on the $\tau_{\rm low\text{-}z}$ budget we separate out in the appendix.

Our analysis is complementary to these. Rather than committing to a particular source or epoch, we leave the redshift $z_x$ of the additional ionization free, adopt a physically motivated, semi-phenomenological burst-plus-recombination-floor history, and compare models through a MCMC on a combined broad dataset. Because the epoch is left free, the data constrain the integrated $\dtau$ far more tightly than $z_x$, so the preferred contribution remains consistent with deposition at redshifts ($z_x\gtrsim 20$) earlier than known astrophysical sources can readily provide. Hence, our analysis encompasses the possibilities of, e.g., exotic high-redshift origins such as accreting NQPBHs~\cite{Qin:2025ymc} or dark-matter energy injection, whereas an analysis focused on a specific scenario might exclude these by construction.

\section{%
\txt{Discussion}%
}\label{sec:disc}%
We have asked, in a deliberately model-agnostic way, what current cosmological
data say about an early, transient contribution to the ionization history. Rather than providing limits, the combined dataset provides a mild $\Delta \chi^2 \sim 6.9-8.3$ or roughly $2 - 2.5\sigma$ preference for the new source. While far short of evidence, it is intriguing, given the other cosmological tensions in the data. Regardless of which data reduction we use, we find preference for a similar $\dtau \simeq 0.012$ and an overall optical depth $\tau\simeq0.073$.
We find the reionization epoch $z_x$ is allowed over a broad range, but with the preferred value dependent on the high-$\ell$ Planck reprocessing we use. We believe the difference likely arises from which $30 < \ell <100$ data are included. 

For PR3, the preference arises as a tension between the value of $\tau$ desired by the low-$\ell$ and high-$\ell$ data. SRoll2, on its own, has no preference for the burst. In contrast, with PR4
the largest contributor to the preference is LoLLiPoP, contributing about half of the total $\Delta\chi^2$, and which appears to prefer a burst on its own. Additional data do strengthen it further, although there is a notable, mild tension in the higher-$\ell$ {\it Planck} data. 

Previous groups have suggested a connection between an enhanced $\tau$ and the current cosmological
preference for a sub-zero effective neutrino mass. Using the same joint
dataset, and allowing for the presence of an additional burst (but not requiring an enhanced $\tau$), we find that allowing $\Sigma m_\nu$ to float 
measurably shifts its posterior toward less negative values relative to
$\Lambda$CDM, without resolving the tension outright.

Upcoming CMB experiments test this scenario, but the prospects depend
sharply on where $z_x$ actually sits. Unlike a fixed-shape optical-depth
increase, an early burst leaves a redshift-dependent signature in
the CMB polarization spectrum. Its characteristic angular scale shifts as
the epoch $z_x$ changes, moving toward the acoustic-peak region at higher
redshift (visible in Figs.~\ref{fig:pr3residual},~\ref{fig:hilresidual_camspec} ~\ref{fig:hilresidual_hillipop}). 
The Cosmology Large Angular Scale Surveyor (CLASS) is a ground-based telescope array that has measured $E$-mode polarization and constrained the optical depth by cross-correlating with {\it Planck} data~\cite{CLASS:2025khf}---future improvements in their map depth and data analysis can deliver constraints on the optical depth independent of {\it Planck}.
LiteBIRD's
dedicated low-$\ell$ polarization program~\cite{LiteBIRD:2022cnt} is also well
positioned to sharpen or definitively test this signature at lower
redshift, given its low noise and independent systematics relative to
{\it Planck}. 
Simons Observatory (SO)~\cite{SimonsObservatory:2018koc}, specifically the Large Aperture Telescope (LAT)~\cite{SimonsObservatory:2025wwn}, is not
expected to improve on {\it Planck}'s own $EE$ sensitivity over $\ell\sim30$--$100$
specifically.
SO LAT's strength lies at higher
$\ell$, where our best-fit models predict non-zero residuals extending into
the acoustic-peak regime and SO's own forecast noise is substantially below
{\it Planck}'s. 
We note, though, that at least part of any such high-$\ell$
signature is expected simply from the shift in standard cosmological
parameters (particularly the amplitude $A_s$) caused by accommodating the additional $\tau$
while remaining consistent with the observed CMB, rather than being a
wholly independent signature of the burst itself; we
have not attempted to cleanly separate the two here. 
On the other hand, the SO Small Aperture Telescopes (SATs)~\cite{SimonsObservatory:2025avm} aim to measure the EE and BB power spectra down to $\ell \sim 50$, and may help probe sources of ionization at high redshifts. As we see in Figs.~\ref{fig:pr3residual},\ref{fig:hilresidual_camspec},\ref{fig:hilresidual_hillipop}, an improvement in this region could be definitive in testing this model vs. \lcdm. 
21\,cm observations of neutral hydrogen also offer a complementary probe here. Because they trace the ionization state of the intergalactic
medium directly at a given redshift, rather than an integrated, degenerate
quantity like $\tau$, they are not limited by how large or small the
CMB-preferred burst amplitude happens to be. One caveat is that computing the 21-cm global signal requires a heating model in conjunction with the ionization history. Nevertheless, a localized increase in $x_e(z)$ would produce a redshift-localized modification of the neutral hydrogen spin temperature, potentially breaking the degeneracy in $z_x$ in the CMB-only analysis. As such, bursts during the dark ages should be visible by next-generation space-based experiments \citep{Burns:2021pkx, Hibbard:2025nov}.

Independently of this redshift-dependent signature, there is a hope to constrain separately the conventional reionization era. Studies such as Ref.~\cite{Kageura:2026ryq} have yielded a result which can be interpreted as $\tau_{\rm low\text{-}z} \simeq 0.055 \pm 0.0075$, and is largely independent of the CMB. Including such a gaussian prior in our fits adds a small additional preference $\sim 1$ to the total $\Delta \chi^2$.   Existing kinematic Sunyaev-Zel'dovich (kSZ) measurements from e.g. ACT also place a constraint on the duration of ordinary, late-time reionization~\cite{Beringue:2025bur,Cain:2025usc}, which is another external check on  $\tau_{\rm low\text{-}z}$, disentangled from any dark-ages contribution. Simons Observatory's own kSZ measurements should improve on this, possibly shrinking the effective $\tau_{\rm low\text{-}z}$ error bar of \cite{Kageura:2026ryq} by a factor of two.

In summary, we have considered a possible new source of reionization, localized at high redshift. We have been motivated by the NQPBH scenario, which would likely yield some additional contribution at high redshift, but we have been agnostic as to the actual source, studying this phenomenologically. Other scenarios such as dark stars or certain forms of dark matter may also yield something similar. While the data are far from definitive, it is intriguing that a mild preference for a new source is present, especially in light of other tensions. The near-term prospect for new data makes it likely that any such source should be definitively tested in the not-so-distant future.

\vskip 0.1in
{\noindent \bf Acknowledgements} 
We thank Miguel Escudero, Daniel Grin, Colin Hill, and Jo Dunkley for useful discussions regarding this work.
NW is supported by NSF under grant PHY-2610067, the BSF under grant 2022287 and the Simons Foundation. NW thanks the TH-group at CERN, where this work was initiated, and M. Escudero, whose conversations prompted this work. 
SK is supported in part by the NSF grant PHY-2609927.
WQ is supported by the Simons Society of Fellows through Grant No. SFI-MPS-SFJ-00006250. 
HAC was supported by the Simons Society of Fellows through Grant No. SFI-MPS-SFJ-00010459.
This work was supported in part through the NYU IT High Performance Computing resources, services, and staff expertise.
The authors acknowledge the Tufts University High Performance Compute Cluster which was utilized for the research reported in this paper.

In this paper, Claude was used to modify \textsc{class}, set up and run MCMCs, and build plots. Claude converted the Cobaya yamls into English descriptions of the parameters of the MCMC. All yamls used for MCMCs whose results appear in this paper have been inspected by humans. Claude was used to find relevant papers and edit the bibliography, which has been verified by humans. Finally, Claude was used to review the text, suggest edits and find corrections, the incorporation or rejection of which was done by humans.

\appendix

\section{Full posteriors}
\label{sec:appendix_full_triangles}
\begin{figure*}[t]
\centering
\includegraphics[width=0.9\textwidth]{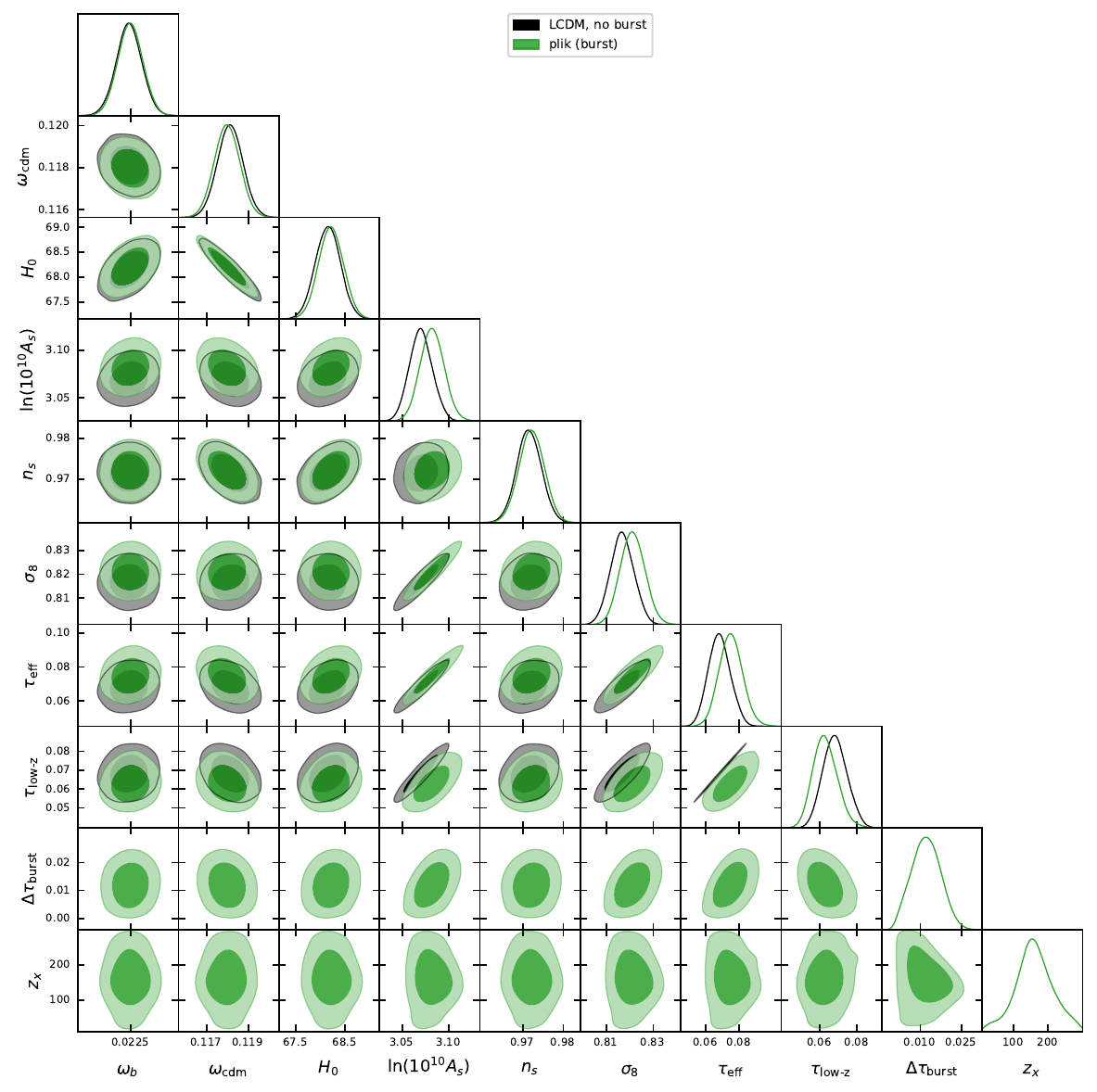}
\caption{Full posterior (base \lcdm\ cosmology plus $\dtau$, $z_x$) for the
burst model using \textsc{plik} (PR3) for the high-$\ell$ likelihood, with
Commander for low-$\ell$ $TT$ and SRoll2 for low-$\ell$ $EE$, vs \lcdm. The
optical depth plotted is $\tau_{\rm eff}$ of Eq.~\eqref{eq:taueff}, and
$\tau_{\rm low\text{-}z}\equiv\tau_{\rm eff}-\dtau$; see
Appendix~\ref{sec:tau_definition}.}
\label{fig:full_triangle_plik}
\end{figure*}

\begin{figure*}[t]
\centering
\includegraphics[width=0.9\textwidth]{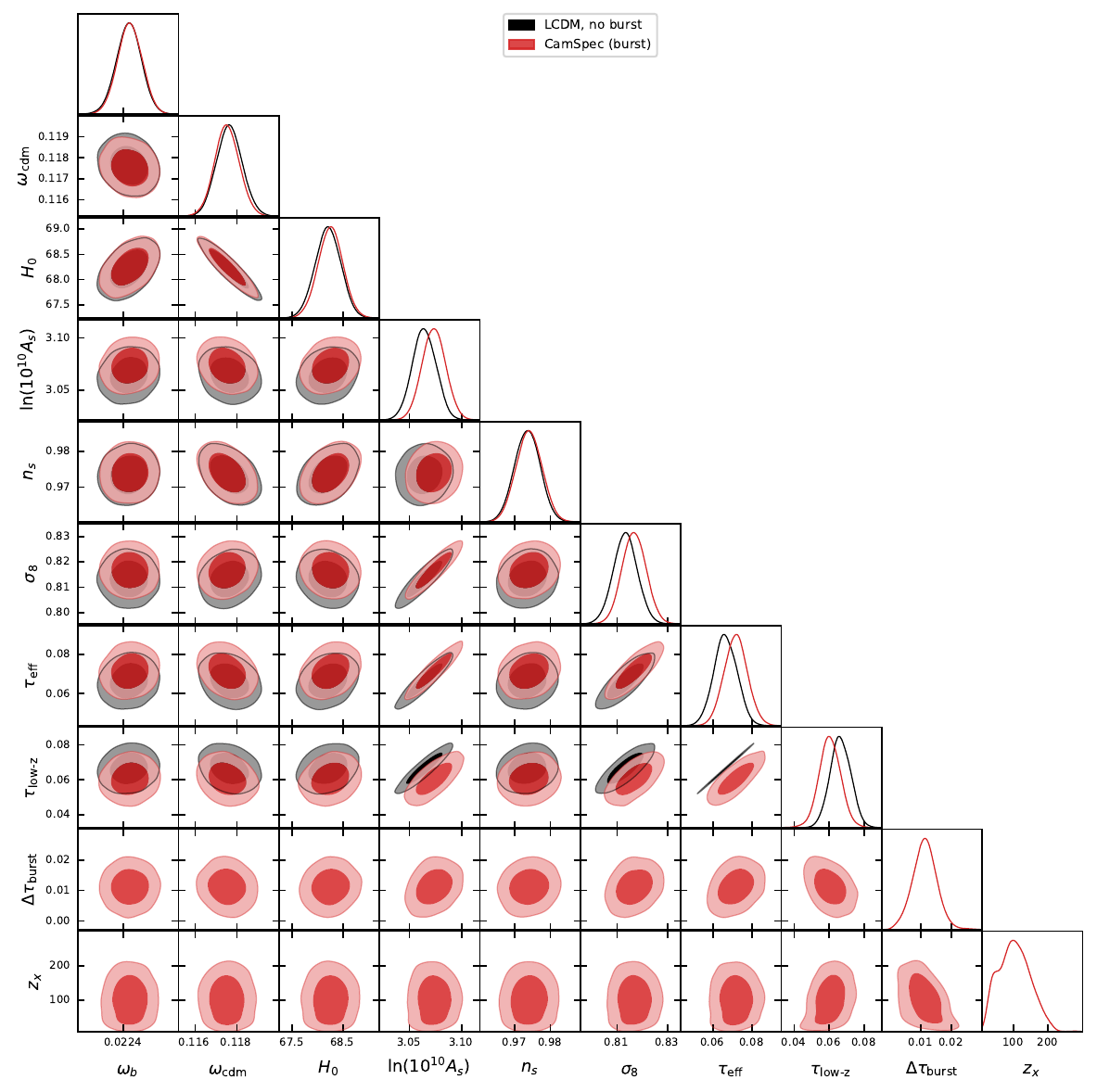}
\caption{Same as Fig.~\ref{fig:full_triangle_plik}, using CamSpec instead
for the high-$\ell$ likelihood and \textsc{LoLLiPoP} for low-$\ell$ $EE$.}
\label{fig:full_triangle_camspec}
\end{figure*}

\begin{figure*}[t]
\centering
\includegraphics[width=0.9\textwidth]{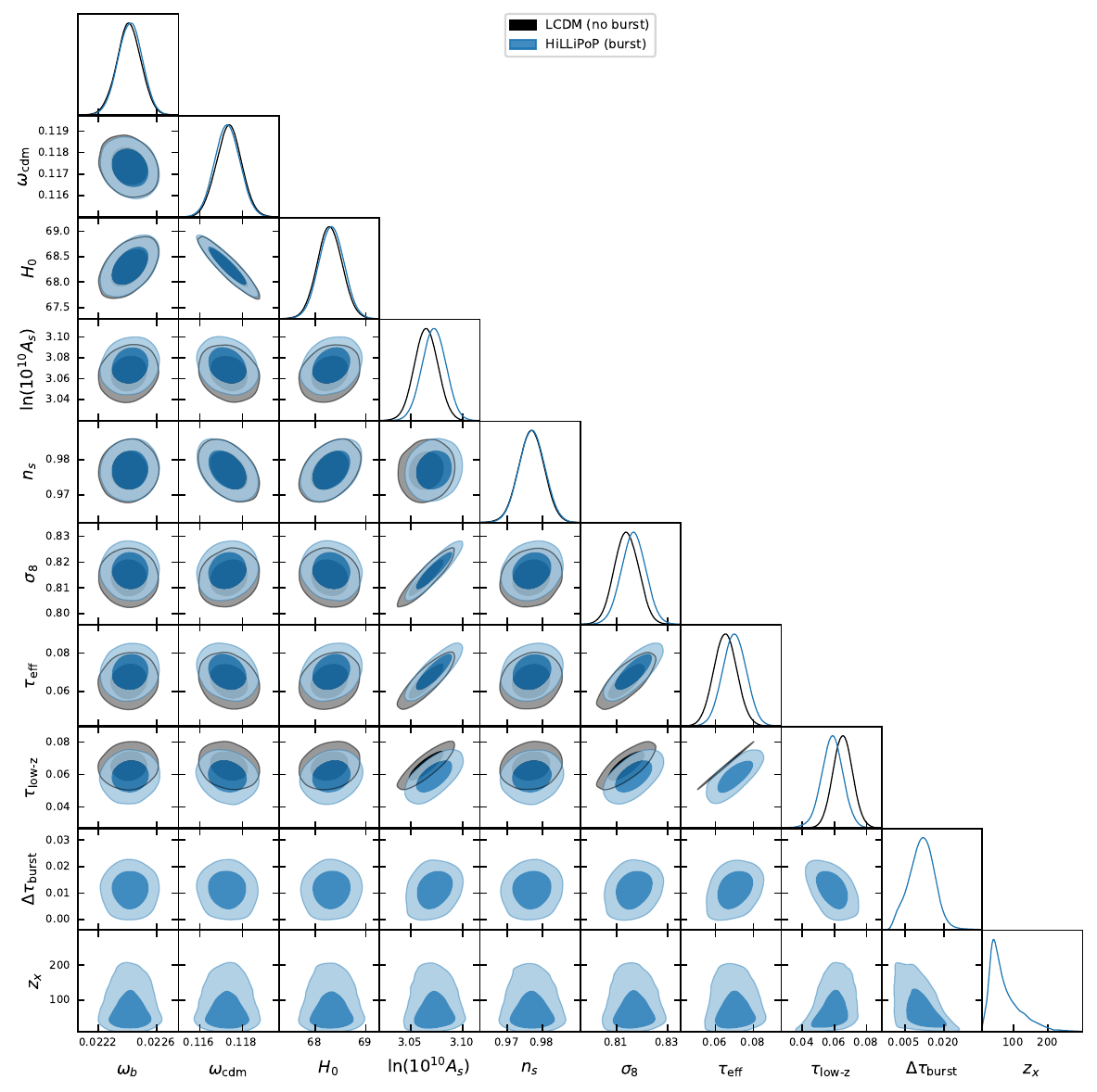}
\caption{Same as Fig.~\ref{fig:full_triangle_camspec}, using HiLLiPoP
instead for the high-$\ell$ likelihood.}
\label{fig:full_triangle_hillipop}
\end{figure*}

Fig.~\ref{fig:triangle_overlay} in the main text shows only the parameters
most relevant to the burst itself. Figs.~\ref{fig:full_triangle_plik},
\ref{fig:full_triangle_camspec} and~\ref{fig:full_triangle_hillipop} show the full
posterior, base cosmology included, for \textsc{plik}, CamSpec, and HiLLiPoP
separately, each against its own matched \lcdm\ (no-burst) control.

\section{The definition of $\tau$ in the presence of a burst}
\label{sec:tau_definition}
\textsc{class} does not integrate the reionization optical depth over a fixed
redshift range. Its \texttt{thermodynamics\_reionization\_get\_tau} routine
locates the global minimum of $\xe(z)$ at some redshift $z_\mathrm{min}$ and integrates 
the Thomson scattering rate~\cite{Greene:2023cro},
\begin{equation}
    \dot{\kappa} = a n_e \sigma_T
\end{equation}
from $z=0$ up to that redshift only, where here $a$ represents the scale factor and $\sigma_T$ is the Thomson cross-section. Whatever lies above the minimum is excluded from
the reported $\tau_{\rm reio}$.
In other words, \textsc{class} defines the optical depth using
\begin{equation}
    \tau_{\rm reio} = \kappa(0),\end{equation} where
    \begin{equation}
    \kappa(z)=\int^{z_\mathrm{min}}_z dz\frac{\dot{\kappa}}{H(z)} .
\end{equation}

For ordinary reionization this is unambiguous: $\xe$ falls to a relic plateau
and rises again, and the minimum sits just below the reionization epoch. A
burst changes this. If the burst has a high enough amplitude such that the new frozen-out value of $\xe$ (see Eq.~\eqref{eq:freeze}) is higher than the previous plateau, the
minimum jumps to a redshift \emph{above} $z_x$ and the burst is counted in
$\tau_{\rm reio}$; if not, the minimum stays at $z\simeq12$ and the burst is
not counted at all in $\tau$. For our favored values of $\dtau\sim 0.01$, we find that around $z_x\sim220$, there is an abrupt shift where the burst (by this mechanism) is no longer included in $\tau_{\rm reio}$. The actual optical depth does not change abruptly, only what is included in $\tau_{\rm reio}$.

Because $\tau_{\rm reio}$ is a sampled input, \textsc{class} solves for the
$z_{\rm reio}$ in both cases where $\dtau$ is included and not. Below the $z_x\sim 220$ boundary, the burst is included, above it generally is not.

We therefore do not use the reported $\tau_{\rm reio}$ for our plots. We integrate the optical depth
ourselves to a fixed $z_{\rm ref}=400$, safely above the $z_x$ prior, and
subtract the same integral for the identical cosmology with both reionization
and the burst switched off, which removes the residual post-recombination
scattering common to every model:
\begin{equation}
  \tau_{\rm eff} \;=\; \big[\kappa(0) - \kappa(z_{\rm ref})\big]_{\rm model}
                 \;-\; \big[\kappa(0) - \kappa(z_{\rm ref})\big]_{\rm no\ reio}.
  \label{eq:taueff}
\end{equation}
This is continuous in $z_x$ by construction, always inclusive of the burst, and
agrees with $\tau_{\rm reio}$  for \lcdm. It is also
insensitive to the precise choice of $z_{\rm ref}$.

It is worth noting that this sampling is far from optimal. Because $\tau_{\rm reio}$ is the sampled input
and its meaning changes at the transition, the flat $\mathcal{U}(0.02,0.12)$
prior of Table~\ref{tab:sampled} maps to a discontinuously different set of
physical total optical depths on either side of it. Future burst studies would be well served to use a variable such as $\tau_{\rm eff}$ in the sampling.

\section{A Description of the MCMC}
\label{sec:mcmc_appendix}
We wish to sample a ``new'' contribution to reionization ($\dtau$) in addition to conventional sources. Since we wish to sample over actual changes to $\tau$, it is inconvenient to use $\ax$. 
In particular, a given contribution in $\dtau$ can be obtained from a smaller value of $\ax$ at a higher redshift.
Reasonable ranges of $\tau$ then imply a very significant prior volume weight on lower redshifts, assuming a flat prior in $A_x$. A complication is that one cannot cleanly separate the ``new'' contributions and the ``conventional'' contributions since the additional source can leak into late times and, in addition, the elevated floor in $\xe$ contributes to the total optical depth, as well.

In \textsc{class}, $\tau_{\rm reio}$ is the sampled base-cosmology
parameter (Table~\ref{tab:sampled}). It is \emph{not} the total optical depth:
\textsc{class} integrates it only up to the global minimum of $\xe(z)$, so with
a burst present it may or may not contain one; see
Appendix~\ref{sec:tau_definition}. \textsc{class} finds it by adjusting the
standard reionization redshift $z_{\rm reio}$ until the resulting
$\tau_{\rm reio}$ matches the sampled value. In our search, this occurs with the source left on
throughout that search. Thus,  \textsc{class} compensates by
adjusting
$z_{\rm reio}$, e.g., part of the burst's own optical depth gets silently reassigned to ``standard''
reionization, and the two sources are no longer separable in the sampled
$\tau_{\rm reio}$ alone.

We fix this with two calculations at the same $z_{\rm reio}$ and
identical values of every other cosmological parameter, one
with the exotic source of ionization and one without. Their optical-depth difference is
$\dtau$: the source's full contribution, including the small
residual ionization excess visible in Fig.~\ref{fig:xebumps} between the
active burst epoch and $z_{\rm reio}$.
Since the sourced
ionization feeds back into HyRec's nonlinear rate equations rather
than being superposed afterward, $\dtau(\ax,\zx)$ has no closed form and is
 nonlinear in $\ax$.
We precompute it on a grid of $(\ax,\zx)$ before the MCMC runs, one pair
of the two calculations above per grid point, and invert the table during
the MCMC to get the $\ax$ corresponding to whatever $(\dtau,\zx)$ is
currently sampled. 

The table is built at one fixed fiducial cosmology
($\omega_b=0.02237$, $\omega_{\rm cdm}=0.12$, $H_0=67.4$, $n_s=0.9649$,
$\ln(10^{10}A_s)=3.044$, $z_{\rm reio}=8$) and does not track how
$\dtau(A_x,\zx)$ depends on the other cosmological parameters, which do
float in the MCMC. However, that only affects the precise labeling of $\dtau$: every likelihood evaluation still runs the exact, sourced \textsc{class}
calculation at the actual sampled cosmology with the derived $A_x$, so both the
fit to data and whether a nonzero source is preferred, comes from actual physics, not the table. 
We have checked that when varying the cosmological parameters within their priors, the resulting $\dtau$ only changes by 1\%.
Hence, the number we quote for $\dtau$ shifts very
slightly, but the statistical preference (or exclusion) itself does not.

The rest of $\tau_{\rm eff}$ is the conventional component at low redshift,
\begin{equation}
\tau_{\rm low\text{-}z} \equiv \tau_{\rm eff} - \dtau.
\label{eq:taulowz}
\end{equation}
Here $\tau_{\rm eff}$ from Eq.~\eqref{eq:taueff} is used rather than \textsc{class}'s
sampled $\tau_{\rm reio}$. The two agree wherever the burst falls inside
\textsc{class}'s integration range, which is most of the posterior, but not
above the $z_x\sim 220$ transition discussed in Appendix~\ref{sec:tau_definition}, where using
$\tau_{\rm reio}$ here introduces a pathology by subtracting a burst that was never included. There are
significant arguments that there is some minimal contribution to $\tau$ at
low redshift (see, e.g., \cite{Kageura:2026ryq}). We note this constraint, but we do not
impose it as a cut. We sample  $\zx \in [10,300]$.

\subsection{Sampled and fixed parameters}
\label{sec:params}

We sample the parameter sets listed in Table~\ref{tab:sampled}: 21 dimensions
for the CamSpec run, 29 for HiLLiPoP (both share the same 13-dimensional base
cosmology, burst parameters, and common-calibration block; they differ in which
high-$\ell$ residual/foreground nuisances are floated, since CamSpec and
HiLLiPoP do not share a common nuisance model). This floats every
foreground/residual nuisance parameter each likelihood declares.

\begin{table*}[t]
\centering
\small
\begin{tabular}{ll@{\hspace{1.2em}}l@{\hspace{1.2em}}l}
\hline
\hline
Parameter & \multicolumn{3}{l}{Prior} \\
\hline
\hline
\multicolumn{4}{l}{\textbf{Base cosmology (6), shared}} \\
$\omega_b$ & \multicolumn{3}{c}{$\mathcal{U}(0.019,0.025)$} \\
$\omega_c$ & \multicolumn{3}{c}{$\mathcal{U}(0.1,0.14)$} \\
$H_0$ & \multicolumn{3}{c}{$\mathcal{U}(55,85)$} \\
$\ln(10^{10}A_s)$ & \multicolumn{3}{c}{$\mathcal{U}(2.9,3.2)$} \\
$n_s$ & \multicolumn{3}{c}{$\mathcal{U}(0.92,1.01)$} \\
$\tau_{\rm reio}$ & \multicolumn{3}{c}{$\mathcal{U}(0.02,0.12)$} \\
\hline
\multicolumn{4}{l}{\textbf{Burst (2), shared}} \\
$\dtau$ & \multicolumn{3}{c}{$\mathcal{U}(0,0.12)$} \\
$\zx$ & \multicolumn{3}{c}{$\mathcal{U}(10,300)$} \\
\hline
\multicolumn{4}{l}{\textbf{Common calibration (5), shared}} \\
$A_{\rm planck}$ & \multicolumn{3}{c}{$\mathcal{N}(1.0,0.0025)$} \\
$A_{\rm ACT}$ & \multicolumn{3}{c}{$\mathcal{U}(0.5,1.5)$} \\
$P_{\rm ACT}$ & \multicolumn{3}{c}{$\mathcal{U}(0.9,1.1)$} \\
$T_{\rm cal}^{\rm SPT}$ & \multicolumn{3}{c}{$\mathcal{U}(0.8,1.2)$} \\
$E_{\rm cal}^{\rm SPT}$ & \multicolumn{3}{c}{$\mathcal{U}(0.8,1.2)$} \\
\hline
\hline
\multicolumn{4}{l}{\textbf{CamSpec's own residual nuisances (8)}} \\
$\rm amp_{143}$, $\rm amp_{217}$, $\rm amp_{143\times217}$ & \multicolumn{3}{l}{$\mathcal{U}(0,50)$ each} \\
$n_{143}$, $n_{217}$, $n_{143\times217}$ & \multicolumn{3}{l}{$\mathcal{U}(0,5)$ each} \\
$\rm cal_{TE}$, $\rm cal_{EE}$ & \multicolumn{3}{l}{$\mathcal{N}(1,0.01)$ each} \\
\hline
\multicolumn{4}{l}{\textbf{HiLLiPoP's own foreground and calibration nuisances (16)}} \\
$A_{\rm radio}$, $A_{\rm dusty}$ & \multicolumn{3}{l}{$\mathcal{U}(0,150)$, $\mathcal{U}(0,100)$} \\
$A_{\rm dustT}$, $\beta_{\rm dustT}$ & \multicolumn{3}{l}{$\mathcal{N}(1.0,0.1)$, $\mathcal{N}(1.51,0.01)$} \\
$A_{\rm dustP}$, $\beta_{\rm dustP}$ & \multicolumn{3}{l}{$\mathcal{N}(1.0,0.1)$, $\mathcal{N}(1.59,0.02)$} \\
$A_{\rm CIB}$, $\beta_{\rm CIB}$ & \multicolumn{3}{l}{$\mathcal{U}(0,20)$, $\mathcal{N}(1.75,0.06)$} \\
$A_{\rm tSZ}$, $A_{\rm kSZ}$, $\xi$ & \multicolumn{3}{l}{$\mathcal{U}(0,50)$, $\mathcal{U}(0,50)$, $\mathcal{N}(0.1,0.1)$} \\
$c_{100A}, c_{100B}, c_{143B}, c_{217A}, c_{217B}$ & \multicolumn{3}{l}{$\mathcal{U}(0.9,1.1)$ each} \\
\hline
\multicolumn{4}{l}{\textbf{\textsc{plik}'s own foreground nuisances (20)}} \\
$\rm calib_{100T}$, $\rm calib_{217T}$ & \multicolumn{3}{l}{$\mathcal{N}(1.0002,0.0007)$, $\mathcal{N}(0.99805,0.00065)$} \\
$A^{\rm CIB}_{217}$, $\xi^{\rm tSZ\times CIB}$ & \multicolumn{3}{l}{$\mathcal{U}(0,200)$, $\mathcal{U}(0,1)$} \\
$A^{\rm tSZ}_{143}$, $A^{\rm kSZ}$ & \multicolumn{3}{l}{$\mathcal{U}(0,10)$ each} \\
$A^{\rm dust\,TT}_{100}$, $A^{\rm dust\,TT}_{143}$, $A^{\rm dust\,TT}_{143\times217}$, $A^{\rm dust\,TT}_{217}$ & \multicolumn{3}{l}{$\mathcal{N}(8.6,2)$, $\mathcal{N}(10.6,2)$, $\mathcal{N}(23.5,8.5)$, $\mathcal{N}(91.9,20)$} \\
$A^{\rm PS}_{100}$, $A^{\rm PS}_{143}$, $A^{\rm PS}_{143\times217}$, $A^{\rm PS}_{217}$ & \multicolumn{3}{l}{$\mathcal{U}(0,400)$ each} \\
$A^{\rm dust\,TE}_{100}$, $A^{\rm dust\,TE}_{100\times143}$, $A^{\rm dust\,TE}_{100\times217}$ & \multicolumn{3}{l}{$\mathcal{N}(0.13,0.042)$, $\mathcal{N}(0.13,0.036)$, $\mathcal{N}(0.46,0.09)$} \\
$A^{\rm dust\,TE}_{143}$, $A^{\rm dust\,TE}_{143\times217}$, $A^{\rm dust\,TE}_{217}$ & \multicolumn{3}{l}{$\mathcal{N}(0.207,0.072)$, $\mathcal{N}(0.69,0.09)$, $\mathcal{N}(1.938,0.54)$} \\
\hline
\end{tabular}
\caption{The sampled (floated) parameters for the three high-$\ell$
likelihood choices discussed in Sec.~\ref{sec:full_suite} -- 21 dimensions
for CamSpec, 29 for HiLLiPoP and 33 for \textsc{plik} (13 shared, plus 8
CamSpec-only, 16 HiLLiPoP-only, or 20 \textsc{plik}-only
residual/foreground/calibration nuisances). $\mathcal{U}$ =
uniform, $\mathcal{N}$ = Gaussian (mean, standard deviation). The
\textsc{plik} nuisances are declared by the likelihood itself rather than in
our configuration, and take its published Planck 2018 priors; \textsc{plik}
additionally carries the joint $A^{\rm tSZ}$--$A^{\rm kSZ}$ prior supplied with
that likelihood, which is included in every prior-inclusive $\Delta\chi^2$ we
quote.}
\label{tab:sampled}
\end{table*}

\begin{table*}[t]
\centering
\begin{tabular*}{\textwidth}{@{\extracolsep{\fill}}llll}
\hline
 & \multicolumn{2}{c}{PR4 (NPIPE)} & PR3 \\
Parameter & CamSpec & HiLLiPoP & \textsc{plik} \\
\hline
\multicolumn{4}{l}{\textbf{\lcdm\ (no burst)}} \\
$\omega_b$        & $0.02245$  & $0.02239$  & $0.02248$ \\
$\omega_{\rm cdm}$ & $0.11770$  & $0.11756$  & $0.11794$ \\
$H_0$              & $68.19$    & $68.19$    & $68.22$ \\
$n_s$              & $0.9745$   & $0.9775$   & $0.9736$ \\
$\tau_{\rm eff}$  & $0.0663$   & $0.0642$   & $0.0698$ \\
$\tau_{\rm eff}$ (95\% range) & $[0.0549, 0.0781]$ & $[0.0535, 0.0772]$ & $[0.0559, 0.0812]$ \\
$\chi^2$            & $6143.12$  & $5617.60$  & $4427.90$ \\
\hline
\hline
\multicolumn{4}{l}{\textbf{Burst}} \\
$\omega_b$        & $0.02245$  & $0.02242$  & $0.02252$ \\
$\omega_{\rm cdm}$ & $0.11756$  & $0.11731$  & $0.11809$ \\
$H_0$              & $68.24$    & $68.32$    & $68.21$ \\
$n_s$              & $0.9741$   & $0.9785$   & $0.9732$ \\
$\tau_{\rm eff}$  & $0.0735$   & $0.0713$   & $0.0760$ \\
$\tau_{\rm eff}$ (95\% range) & $[0.0600, 0.0836]$ & $[0.0580, 0.0821]$ & $[0.0619, 0.0891]$ \\
$\dtau$            & $0.0123$   & $0.0135$   & $0.0133$ \\
$\dtau$ (95\% range) & $[0.0033, 0.0192]$ & $[0.0011, 0.0199]$ & $[0.0017, 0.0218]$ \\
$z_x$              & $106$      & $47.8$     & $165$ \\
$z_x$ (95\% range) & $[19.4, 186.5]$ & $[10.0, 168.3]$ & $[44.3, 267.2]$ \\
$\chi^2$            & $6135.07$  & $5610.74$  & $4421.81$ \\
\hline
$\Delta\chi^2$ (data only) & $+8.05$  & $+6.85$ & $+6.10$ \\
$\Delta\chi^2$ (incl.\ prior) & $+8.33$ & $+7.29$ & $+6.90$ \\
\hline
\end{tabular*}
\caption{
Best-fit parameters for each dataset choice, all at fixed
$N_{\rm eff}=3.044$ and $\mnu=0.059\ev$.}
\label{tab:bestfit}
\end{table*}

The best-fit values reported throughout (Table~\ref{tab:bestfit}) are
refined beyond the single best-sampled
chain point, which is only as reliable as how densely that particular
region of parameter space happens to have been explored. We first tried a
standard derivative-free optimizer (BOBYQA) seeded at the chain's own best
point, but found it unreliable on these high-dimensional, multi-likelihood
posteriors. We instead refine each best fit iteratively: shrink the
sampled prior range around the current best point and cool the MCMC's own
sampling temperature (sharpening the posterior toward its mode) over
several rounds, always re-centering on whichever point has proven deepest
so far, so the reported $\chi^2$ can only improve, never worsen, relative
to the chain's own best sample. We use this both for the global best fit
and for the drop-set points in Table~\ref{tab:dropsets}, as well as for the profile likelihood in Fig.~\ref{fig:zx_profile}.

\begin{figure*}[htbp]
\centering
\eqd{\includegraphics[width=\textwidth]{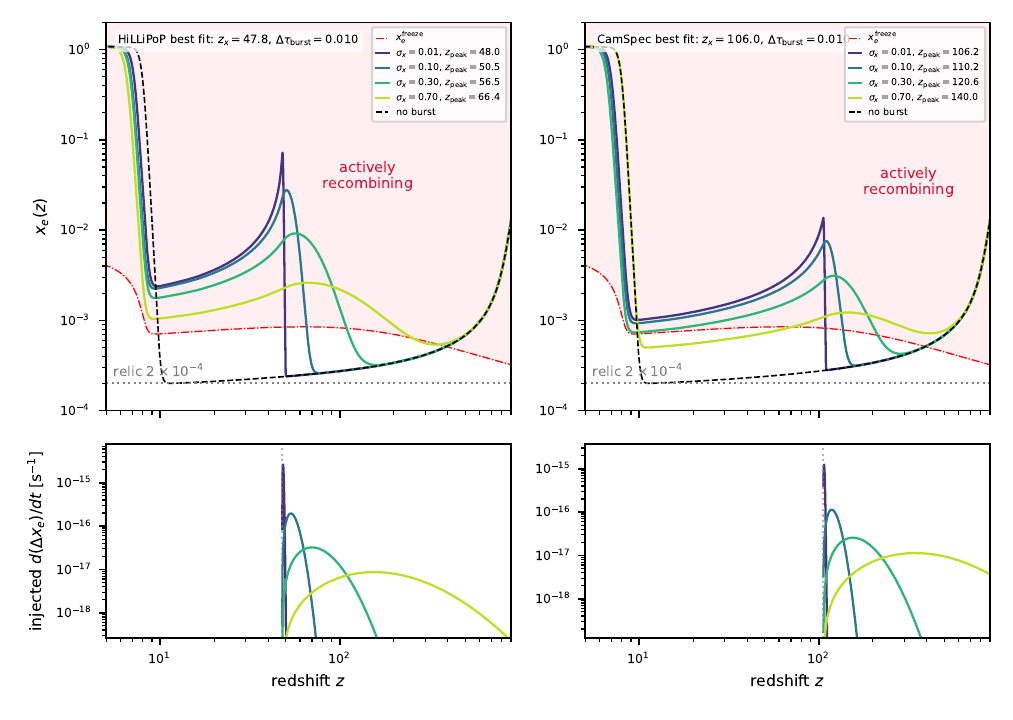}}%
\caption{Same style as Fig.~\ref{fig:xebumps}, but varying the burst width
$\sigma_x=0.01,0.1,0.3,0.7$ at fixed total $\dtau=0.010$ (same normalization
convention as Fig.~\ref{fig:xebumps}) for two representative $z_x$:
HiLLiPoP's own best fit ($z_x=47.8$, left) and CamSpec's ($z_x=106$, right). Top: $x_e(z)$; bottom: the injected
rate $d(\Delta x_e)/dt$.}
\label{fig:xebumps_sigma}
\end{figure*}

Fixed parameters fall into two categories. First, the physical model
itself: $N_{\rm eff}=3.044$ (via $N_{\rm ur}=2.0308$, one massive species)
and $\mnu=0.059$~eV (the minimal normal-ordering value), together with the
burst width $\sx=0.1$. As discussed in the main text, for $\sigma_x\lesssim 0.2$, the model is well described as a localized ionizing period, and the fit does not significantly depend on it. For larger values, this is no longer true, and the source, especially including the subsequent recombination, is no longer localized. We see this in Fig.~\ref{fig:xebumps_sigma}.

HiLLiPoP fixes several parameters internally regardless
of any user configuration: one per-frequency calibration factor
($c_{143A}=1.0$, the pivot for the five relative calibrations in
Table~\ref{tab:sampled}), all six polarization efficiencies, the radio
spectral index ($\beta_{\rm radio}=-0.8$), and both synchrotron amplitudes
(zero). The CIB, dust, and Sunyaev--Zel'dovich foreground parameters
listed in Table~\ref{tab:sampled} are floated in the analysis
used throughout this paper.

\subsection{A Visual Comparison}
\label{sec:rebinning}

\begin{figure}[t]
\centering
\includegraphics[width=0.48\textwidth]{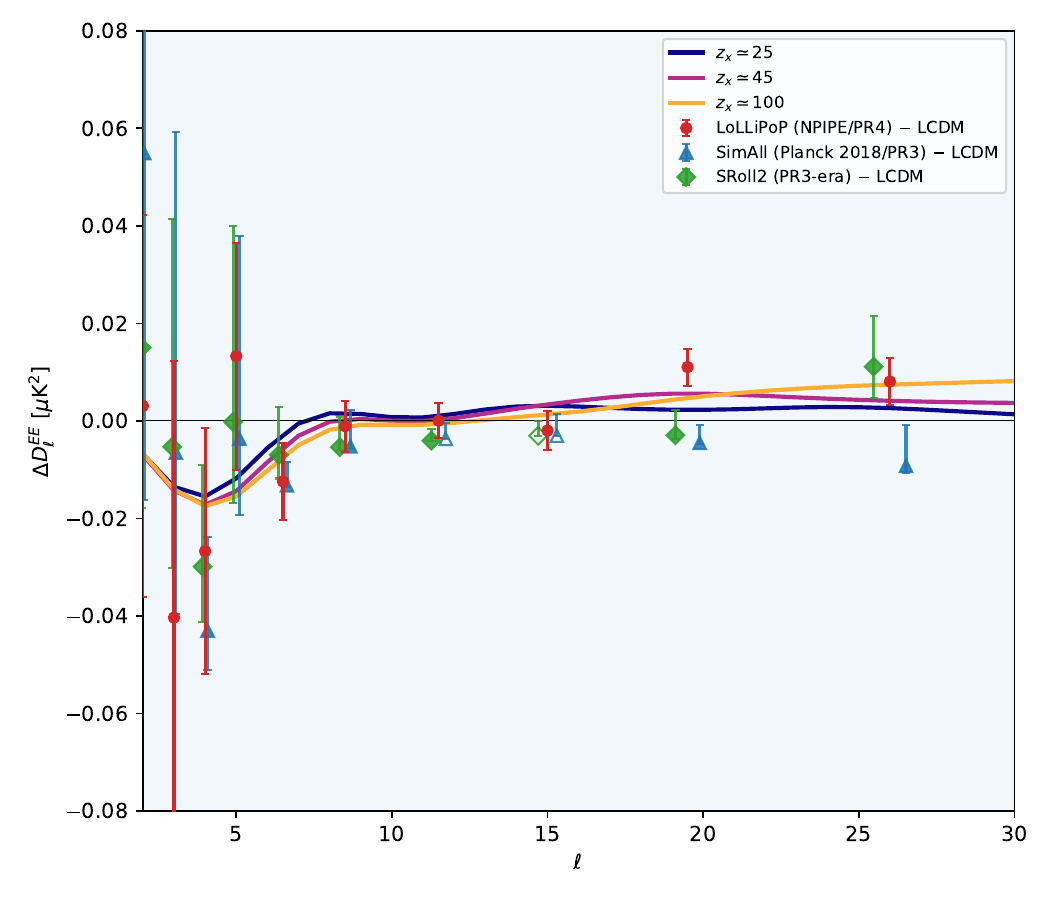}
\caption{Low-$\ell$ $EE$ residual (burst $-$ $\Lambda$CDM) vs.\ $\ell$ for
the same minimized best-fit points as Fig.~\ref{fig:hilresidual_hillipop}
($z_x=25,45,100$), against each of the three reprocessings'
own data residual, all rebinned onto the same log-$\ell$ bins via
Eq.~\eqref{eq:gls}. Bars are the $\Delta\log P=0.5$ interval; open symbols are
one-sided, the mode being at $C_\ell=0$.}
\label{fig:reprocessing_residual}
\end{figure}

To gain a sense of the tensions between models and data, it is useful to visualize them directly. To this end, we include a number of residual plots to show this, specifically Figs.~
\ref{fig:reprocessing_residual}, \ref{fig:hilresidual_camspec}, \ref{fig:hilresidual_hillipop}, and \ref{fig:pr3residual}. In these plots, we combine
several adjacent data points into one before plotting, purely so
the data can be compared by eye against the best-fit curves without the
point-to-point scatter obscuring the trend. This
rebinning is used only for these plots: the MCMC employs standard likelihoods as described without any additional binning.

The error bars shown are thus a visual aid only. They do not reflect how much of an apparent feature could be reabsorbed by the
many floated nuisance parameters varying within
their own priors, so a by-eye comparison of a curve to these error bars
can over- or under-state the actual significance of any given point. The
$\Delta\chi^2$ values quoted throughout -- computed at the full best-fit
point, nuisance parameters included -- are the correct summary of the
fit's significance. These plots are for visual comparison only and enter
none of the reported calculations.

Given a set of points with residuals $\bm{r}$ (data $-$ theory) and
covariance $C$, we combine them into a single point
\begin{equation}
\hat r = \frac{\mathbf{1}^T C^{-1}\bm{r}}{\mathbf{1}^T C^{-1}\mathbf{1}},
\qquad
\sigma_{\hat r} = \left(\mathbf{1}^T C^{-1}\mathbf{1}\right)^{-1/2},
\label{eq:gls}
\end{equation}
the inverse-covariance weighted mean and its uncertainty. This holds for any
$C$, diagonal included, in which case it reduces to the familiar
$1/\sigma_i^2$ weighting.

LoLLiPoP, HiLLiPoP, ACT DR6 CMB-only and SPT-3G supply a full $C$, so
Eq.~\eqref{eq:gls} is applied directly to each $\ell$ window's covariance
sub-block.

SimAll and SRoll2 (Fig.~\ref{fig:reprocessing_residual}) have no covariance ---
each is a per-multipole profile likelihood, tabulated independently at every
$\ell$. We take the mode of each table as the point and weight by the symmetrised
$\Delta\log P=0.5$ width, but combine the interval's two sides separately rather
than symmetrising them. This is approximate for a non-Gaussian likelihood, but
adequate for the by-eye comparison in Fig.~\ref{fig:reprocessing_residual}.

\subsection{Dependence on low-$\ell$ $EE$ reprocessing}

\begin{figure}[htbp]
\centering
\includegraphics[width=0.48\textwidth]{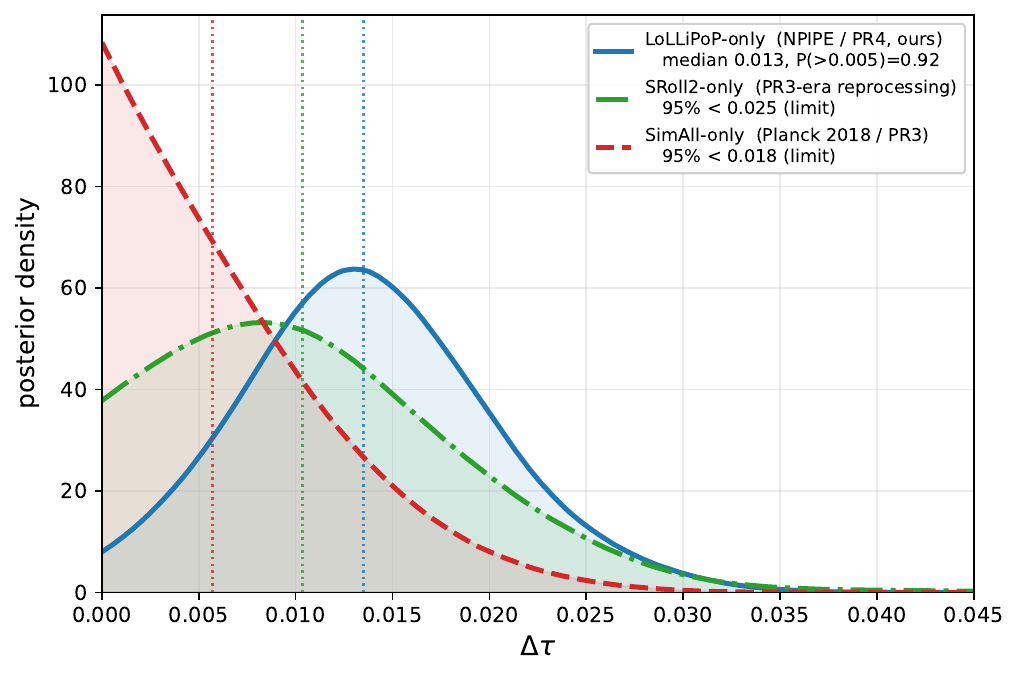}
\caption{Posterior on $\dtau$ from the low-$\ell$ $EE$ likelihood alone, for
three reprocessings of the same {\it Planck} data: \textsc{LoLLiPoP} (PR4),
SRoll2 (PR3) and \textsc{SimAll} (PR3). Dotted lines mark each median; $z_x$ is
restricted to $[50,500]$.}
\label{fig:reprocessing}
\end{figure}

Since  much of the preference for $\dtau$ traces to the low-$\ell$
$EE$ polarization likelihood (Sec.~\ref{sec:lollipop_alone}), we check whether it
depends on which reprocessing of that same {\it Planck} low-$\ell$ polarization
data is used. Fig.~\ref{fig:reprocessing} shows the $\dtau$ posterior using
only the low-$\ell$ $EE$ likelihood (no ACT, SPT, HiLLiPoP, BAO, or
supernovae), repeated for three independent pipelines applied to the same
underlying {\it Planck} data: \textsc{LoLLiPoP} (NPIPE/PR4p), \textsc{SimAll} ({\it Planck} 2018/PR3), and
SRoll2 (an independent PR3-era reprocessing). These runs restrict $z_x\in[50,500]$, in order to zoom in on the preferred range we find in our overall analysis. Only LoLLiPoP strongly peaks at nonzero $\dtau$ on
its own (median $0.0135$, $P(\dtau>0.005)=0.92$); the other two are far weaker, with
95\% upper limits of $0.018$ (SimAll) and $0.025$ (SRoll2); SimAll peaks at zero while SRoll2 peaks weakly at $0.010$. 

The upshot of this is somewhat interesting: while SRoll2+\textsc{plik} prefers a burst, it is a tension between the low-$\ell$ and high-$\ell$ data. For NPIPE, LoLLiPoP {\em alone} appears to prefer a burst, with the larger data suite enhancing that preference. 

Fig.~\ref{fig:reprocessing_residual} shows why the three pipelines differ,
at the level of the actual data. We show the
same real minimized best-fit curves from Fig.~\ref{fig:hilresidual_hillipop} against
each pipeline's own low-$\ell$ $EE$ residual, rebinned via Eq.~\eqref{eq:gls} onto the same
log-$\ell$ bins for all three.\footnote{LoLLiPoP is a Gaussian bandpower likelihood
with a full covariance; SimAll and SRoll2 instead report a per-multipole
profile likelihood with no covariance between multipoles, so each native
point's asymmetric $\Delta\ln P=0.5$ interval is carried through the same
rebinning with its two sides combined separately. This is obviously
not exact for a non-Gaussian likelihood, but adequate at this coarseness
for a simple visual comparison.} The changes and the consequences for a burst are clear from this plot. While the reprocessings are all largely consistent in the lower $\ell$ region, in the higher region they are not. Since a significant effect of a burst is to enhance power here, the moderate enhancement of the $25<\ell<30$ region drives a preference for a burst.

\bibliography{refs}
\end{document}